\documentclass[11pt]{article}
\usepackage{epsfig,subfigure}
\usepackage{color}
\usepackage{mathrsfs,amsbsy}
\usepackage{dsfont}
\usepackage{hyperref}
\usepackage[all]{xy}
\usepackage{graphicx}
\usepackage{amsmath,amsfonts,amsthm}
\usepackage{amssymb,latexsym}

\newcommand{\ignore}[1]{}

\def\uinl{\left|\kern-1.5truept\left|\kern-1.5truept\left|}
\def\uinr{\right|\kern-1.5truept\right|\kern-1.5truept\right|}
\def\uinlr{\hbox{$|\kern-1.5truept|\kern-1.5truept|\cdot
                 |\kern-1.5truept|\kern-1.5truept|$}}

\def\loc{{\rm loc}}  
\def\nl{{\rm nl}}

\newcommand{\vm}[1]{\mathbf{#1}}
\newcommand{\bsym}[1]{\boldsymbol{#1}}

\newcommand{\tit}[1]{\textit{#1}}

\newcommand{\mrm}[1]{\mathrm{#1}}
\newcommand{\eps}{\varepsilon}
\newcommand{\eff}{\textrm{eff}}
\newcommand{\epsh}{\widehat\varepsilon}
\newcommand{\psih}{\widehat\psi}

\title{
Hybrid preconditioning
for iterative diagonalization 
of ill-conditioned generalized eigenvalue problems
in electronic structure calculations\thanks{
Supported in part by award no.\ 118128 from the UC Lab Fees Research Program. 
This work performed, in part, under the auspices of the U.S. Department of Energy by Lawrence Livermore National Laboratory under Contract DE-AC52-07NA27344.
} 
}
\author{
Yunfeng Cai\thanks{
Department of Computer Science,
University of California, Davis 95616, USA, 
{\tt yfcai@cs.ucdavis.edu} 
} \quad 
Zhaojun Bai\thanks{Department of 
Computer Science and Department of Mathematics, 
University of California, Davis 95616, USA, 
{\tt bai@cs.ucdavis.edu} 
} \quad  
John E. Pask\thanks{
Condensed Matter and Materials Division, 
Lawrence Livermore National Laboratory, 
Livermore, CA 94550, USA,
{\tt pask1@llnl.gov} 
} \quad
N. Sukumar\thanks{
Department of Civil and Environmental Engineering, 
University of California, Davis 95616, USA, 
{\tt nsukumar@ucdavis.edu} 
} 
}
\date{\today}

\begin{document}
\maketitle


\begin{abstract}
The iterative diagonalization of a sequence of 
large ill-conditioned generalized eigenvalue problems 
is a computational bottleneck in quantum mechanical methods 
employing a nonorthogonal basis 
for {\em ab initio} electronic structure calculations. 
We propose a hybrid preconditioning scheme
to effectively combine global and locally accelerated 
preconditioners for rapid iterative diagonalization of 
such eigenvalue problems. 
In partition-of-unity finite-element (PUFE) 
pseudopotential density-functional calculations, employing a nonorthogonal basis, 
we show that the hybrid preconditioned block steepest
descent method is a cost-effective eigensolver,
outperforming current state-of-the-art global preconditioning schemes, 
and comparably efficient for the ill-conditioned generalized eigenvalue problems produced by PUFE 
as the locally optimal block preconditioned conjugate-gradient method
for the well-conditioned standard eigenvalue problems produced by planewave methods.
\end{abstract}

\section{Introduction}\label{sec:intro}

First principles ({\em ab initio}) quantum mechanical simulations
based on density functional theory (DFT)~\cite{hohe:64,Kohn:65} 
are a vital component of research in condensed matter physics and
molecular quantum chemistry. Using DFT, the many-body
Schr\"{o}dinger equation for the ground state properties 
of an interacting system of electrons and nuclei is reduced to
the self-consistent solution of an effective single-particle 
Schr\"{o}dinger equation, known as the Kohn-Sham equation: 
\begin{equation}\label{eq:ks}
\mathcal{H} \psi_i(r)
= \left[-\frac{1}{2}\nabla^2+V_\eff(r,\rho(r))\right]\psi_i(r)
=\varepsilon_i\psi_i(r),
\end{equation} 
where $\varepsilon_i$ 
are particle energies (eigenvalues) and
$\psi_i$ 
are the associated wavefunctions (eigenfunctions). The
Hamiltonian $\mathcal{H}$ consists of 
kinetic energy operator 
$-\frac{1}{2}\nabla^2$
and effective potential operator $V_\eff(r,\rho(r))$.
The effective potential $V_\eff$ 
depends on the electronic charge density
\begin{equation}\label{density}
\rho(r)=\sum_{i} f_i |\psi_i(r)|^2,
\end{equation}  
where $0 \le f_i \le 2$ is the electronic occupation of state $i$ and
the sum is over all occupied states.
Since $\psi_i$ depends on $V_\eff$ which depends on $\rho(r)$ which depends again on $\psi_i$,
the Kohn-Sham equation \eqref{eq:ks} is a nonlinear eigenvalue problem.

The importance of {\em ab initio} calculations stems from their
underlying quantum-mechanical nature, yielding insights inaccessible to
experiment and robust, predictive power unattainable by more
approximate empirical approaches. However, because \emph{ab initio}
calculations are computationally intensive, a vast range of real
materials problems remain inaccessible by 
such accurate, quantum mechanical means. 
To address this limitation, there has been substantial
effort in recent years to develop 
\emph{ab initio} methods that use efficient, local bases in order to both reduce degrees of freedom and facilitate large-scale parallel implementation:  
augmented planewave plus local orbital (APW+lo)~\cite{sing:05,sjos:00}, atomic-orbital (AO), e.g.,~\cite{arta:08,blum:09}, and 
real-space methods~\cite{beck:00,tors:06,saad:10} such as
finite-difference~\cite{chel:1994a,chel:1994b,BriSulBer95}, wavelet~\cite{cho:1993,aria:99,GenNeeSch08},
finite-element (FE)~\cite{tsu:1995,pask:05},
partition-of-unity finite element (PUFE)~\cite{suku:09,pask:11,pask:11a}, and discontinuous Galerkin (DG)~\cite{lin:2012a} methods, among many others, 
see for example~\cite{mart:04}. 

In the vast majority of \emph{ab initio} methods, 
the dominant computational cost is 
the iterative diagonalization of the sequence 
of large linear eigenvalue problems  
produced by the discretization of equation~\eqref{eq:ks}  
in the chosen basis~\cite{saad:96,kress:96,yang:05,saad:06,saad:10,zhou:06,vome:08,rays:08}. 
The linear eigenvalue problems produced by highly efficient physics based 
APW+lo, AO, and PUFE bases, while smaller than those of other bases, present a particular challenge 
as they are generalized eigenvalue problems 
with ill-conditioned coefficient matrices, and are much more 
difficult to precondition than those produced by conventional 
planewave based methods, due to the lack of diagonal dominance 
and absence of an efficient representation for the inverse Laplacian.

Here, building on prior work \cite{wood:85,sing:89,pask:05a,ande:05,rays:08,blah:10},
we propose a {\em hybrid preconditioning scheme}
for rapid iterative diagonalization of 
the sequence of ill-conditioned generalized Hermitian eigenvalue problems 
produced by modern orbital based electronic structure methods, such as APW+lo, AO, and PUFE. 
The hybrid preconditioning scheme
effectively combines a global shifted-inverse preconditioner as in~\cite{pask:05a,ande:05,blah:10} 
and locally accelerated shifted-inverse preconditioners as in~\cite{wood:85,sing:89,pask:05a,ande:05,rays:08} that
target eigenpairs of interest individually. 
The global preconditioner serves as sole preconditioner in early self-consistent iterations and as convergence accelerator for local preconditioners in subsequent iterations.
We have conducted extensive tests of 
the proposed hybrid preconditioning scheme
with the block steepest descent method
in PUFE pseudopotential density functional calculations
on a variety of systems, including the difficult case of
triclinic metallic CeAl. This system has deep atomic potentials and
15 electrons per unit cell in valence, thus requiring the computation of many, strongly localized eigenfunctions, 
which in turn requires the addition of correspondingly many orbital enrichments in the PUFE electronic structure method. 
Our results reveal that in terms of 
average numbers of inner and outer iterations, 
the hybrid preconditioner performs markedly better than global or local preconditioners alone, 
and the resulting solver performs as well on the ill-conditioned generalized eigenvalue problems produced by the PUFE \tit{ab initio} method as does
the locally optimal block preconditioned conjugate-gradient (LOBPCG) method 
on well-conditioned \tit{standard} eigenvalue problems produced by the planewave method. 

The remainder of the paper is organized as follows. 
In Section~\ref{sec:scf}, we outline 
the self-consistent field (SCF) procedure and 
iterative diagonalization process in an algebraic
setting, and discuss the ill-conditioned 
generalized eigenvalue problems produced by the PUFE electronic structure method. 
In Section~\ref{sec:labpsd}, 
we describe the hybrid preconditioning scheme 
and its use in the block steepest descent method. 
Implementation details  
are presented in Section~\ref{sec:impl}. 
Numerical results are presented in
Section~\ref{sec:num} and we close with final remarks
in Section~\ref{sec:conclusion}.

\section{SCF, iterative diagonalization, 
and ill-conditioned GHEPs} \label{sec:scf}

Electronic structure methods such as APW+lo, AO, and PUFE methods incorporate information from local atomic solutions to construct efficient bases for molecular or condensed matter calculations. 
This information is typically incorporated in the form of localized, atomic-like basis functions (orbitals), 
which generally leads to a nonorthogonal basis. Discretization of the Kohn-Sham equation \eqref{eq:ks} in such a basis then leads to a nonlinear algebraic eigenvalue  problem 
\begin{equation}\label{ks-nlep}
{H}(V_\eff) \Psi = S \Psi E,  
\end{equation}
where 
${H}(V_\eff)$ is the discrete KS-Hamiltonian matrix 
and consists of a local part $H^{(\loc)}(V_\eff)$ 
and, when pseudopotentials~\cite{mart:04} are employed, nonlocal part $H^{(\nl)}$:
\[
{H}(V_\eff) = H^{(\loc)}(V_\eff) + H^{(\nl)}.
\] 
$H^{(\loc)}(V_\eff)$ is a Hermitian matrix which 
depends on the effective potential $V_\eff$, 
which in turn depends on the electronic density $\rho(r)$ 
computed from the eigenvectors $\Psi$.
$H^{(\nl)}$ is a low-rank Hermitian matrix associated
with the non-local part of the pseudopotential. 
$S$ is the overlap (Gram) matrix of the basis and is
Hermitian positive-definite.  
The nonlocal matrix $H^{(\nl)}$ and 
overlap matrix $S$ are independent of $V_\eff$, and
hence do not depend on $\rho(r)$ or $\Psi$.
In condensed matter calculations, it is required to sample 
the Brillouin zone~\cite{mart:04} at a sufficient number of $\vm{k}$-points, making the above matrices complex Hermitian rather than real symmetric.
 In addition, for methods whose basis functions are localized, such as wavelet, FE, PUFE, DG, and (to a lesser extent) AO-type methods, the above matrices are sparse: 
for example, in the case of PUFE, having a few hundred nonzero 
entries per row, independent of problem size.

\begin{figure}
\centering
\includegraphics[width=0.80\textwidth]{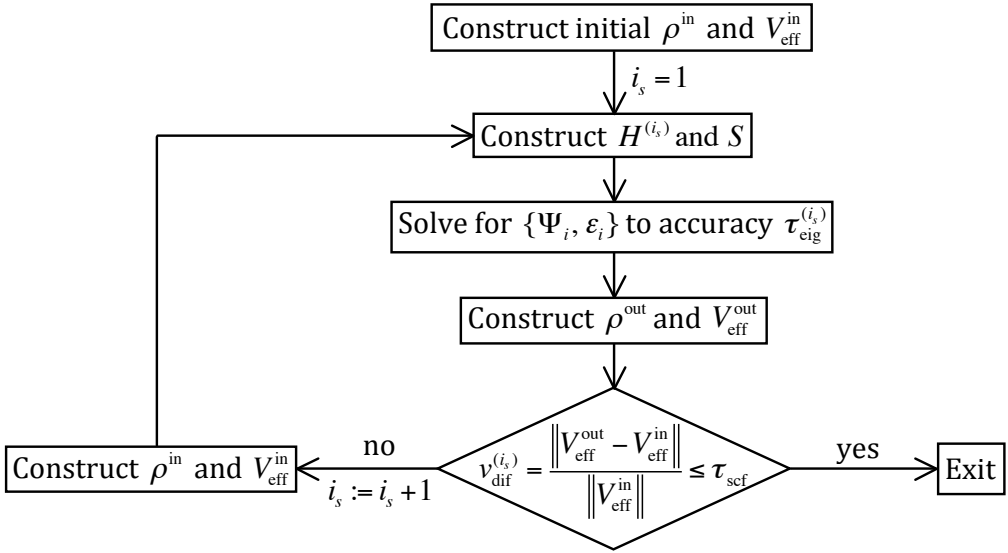}
\caption{Self-consistent field (SCF) procedure.} \label{fig:scf}
\end{figure}

The nonlinear eigenvalue problem~\eqref{ks-nlep} is solved
by fixed-point iteration (see~\cite{mart:04}):
starting with an initial guess for the input charge density 
$\rho^\mrm{in}$ and associated effective potential $V_\eff^\mrm{in}$ and
iterating until the difference between the input and output effective potentials, 
$V^{\rm in}_\eff$ and $V^{\rm out}_\eff$, is within a specified tolerance $\tau_{\rm scf}$; i.e., 
the process is terminated at the $i_s$-th iteration if 
\begin{equation}
\label{eq:vdif}
v_{\rm dif}^{(i_s)}=
\frac{\|V^{\rm out}_\eff-V^{\rm in}_\eff\|}
{\|V^{\rm in}_\eff\|}  \leq \tau_{\rm scf}.
\end{equation}
This process is known as a {\em self-consistent field} (SCF) procedure. 
A schematic of the SCF procedure 
is shown in Figure~\ref{fig:scf}.  

At the $i_s$-th SCF iteration,
with an approximate effective potential $\widetilde{V}_\eff$
extrapolated from 
previous SCF iterations~\cite{pula:80},
the nonlinear eigenvalue problem \eqref{ks-nlep} 
becomes the following linear generalized Hermitian eigenvalue problem (GHEP):
\begin{align} \label{ghep}
H^{(i_s)} \Psi^{(i_s)} = S \Psi^{(i_s)} E^{(i_s)}, 
\end{align}
where 
\begin{align*} 
H^{(i_s)} = H^{(\loc)}(\widetilde{V}_\eff) +H^{(\nl)},
\end{align*}
$H^{(\loc)}(\widetilde V_\eff)$ is a Hermitian matrix, $H^{(\nl)}$ is a low-rank 
Hermitian matrix, $S$ is Hermitian positive definite, and all matrices are sparse when arising from discretization in a localized basis such as PUFE. As the SCF iteration proceeds, changes in $\widetilde{V}_\eff$, 
and thus $H^{(\loc)}(\widetilde{V}_\eff)$, $\Psi^{(i_s)}$, and $E^{(i_s)}$ become smaller and smaller until convergence to the specified tolerance is achieved.

Since in the first few SCF iterations $\widetilde{V}_\eff$ is not yet well converged, 
the GHEP~\eqref{ghep} need not be solved to high accuracy. All that is necessary is that the accuracy be sufficient to
 allow the outer SCF iteration to converge without incurring significant additional iterations relative to exact solution. 
As the SCF iterations proceed and $\widetilde{V}_\eff$ converges, the accuracy requirement for the solution of the GHEP~\eqref{ghep} increases. 
Specifically, from the previous SCF iteration, 
we have an estimate $\{\widehat{E}_0, \widehat{\Psi}_0\}$ 
of the lowest $m$ eigenpairs with the maximum residual norm
\begin{equation} \label{eq:res0} 
\tau_{{\rm eig},0}^{(i_s)}
= \mbox{Res}_{\max}[\widehat{E}_0, \widehat{\Psi}_0],
\end{equation}
where if 
$\widehat{E} = \mbox{diag}(
\widehat{\varepsilon}_1, 
\widehat{\varepsilon}_2, \ldots, 
\widehat{\varepsilon}_m)$
and $\widehat{\Psi} = 
[\widehat{\psi}_1, \widehat{\psi}_2, \ldots, \widehat{\psi}_m]$
are approximate eigenpairs, then
$$\mbox{Res}_{\max}[\widehat{E}, \widehat{\Psi}]
=\max_{1\le i \le m} \mbox{Res}[\widehat{\varepsilon}_i, \widehat{\psi}_i],$$ 
and
$\mbox{Res}[\widehat{\varepsilon}_i, \widehat{\psi}_i]$ is the
relative residual norm
for the approximate eigenpair
$(\widehat{\varepsilon}_i, \widehat{\psi}_i)$ of
GHEP~\eqref{ghep}:
\begin{equation}\label{eq:res1}
\mbox{Res}[\widehat{\varepsilon}_i, \widehat{\psi}_i]  
\equiv \frac{\|r_i\|}{\|H^{(i_s)}\widehat{\psi}_i\|},
\end{equation}
and $r_i = H^{(i_s)} \widehat{\psi}_i - 
\widehat{\varepsilon}_i S \widehat{\psi}_i$.

Our objective at the $i_s$-th SCF iteration 
is to compute the improved estimate 
$\{\widehat{E}_{l}, \widehat{\Psi}_{l}\}$ satisfying 
\begin{equation} \label{eq:resL} 
\mbox{Res}_{\max}[\widehat{E}_{l}, \widehat{\Psi}_{l}] 
\leq \tau_{{\rm eig},l}^{(i_s)}
\end{equation}
where the tolerance $\tau_{{\rm eig},l}^{(i_s)}$ is chosen
to achieve a desired reduction relative to
$\tau_{{\rm eig},0}^{(i_s)}$ and/or ${v_{\rm dif}^{(i_s)}}$.
In practice, one or two orders of magnitude reduction is typically sufficient 
for the SCF procedure to converge in a comparable number 
of iterations to exact solutions (i.e., reduction to zero).\footnote{ 
By backward error analysis~\cite[Chap.5]{bddrv:00},
there exists a matrix 
$\Delta H$ with $\|\Delta H\|=\|r_i\|/{\|\widehat{\psi}_i\|}$
such that $( \widehat{\varepsilon}_i, \widehat{\psi}_i)$ is
an exact eigenpair of the matrix pair $(H^{(i_s)}+\Delta H, S)$.
Consequently, we have
\[ 
\frac{\|\Delta H\|}{\|H^{(i_s)}\|} 
= \frac{ \|r_i\|}{\|H^{(i_s)}\|\|\widehat{\psi}_i\|}\le 
\frac{\|r_i\|}{\|H^{(i_s)}\widehat{\psi}_i\|}.
\]
Therefore,
$\mbox{Res}[\widehat{\varepsilon}_i, \widehat{\psi}_i] \leq tol$
implies relative backward error of 
$(\widehat{\varepsilon}_i, \widehat{\psi}_i)$ 
less than $tol$.
}

Since during the course of the SCF iteration to convergence, a wide range of accuracies are required for the solution of the GHEP~\eqref{ghep} 
and excellent approximations are available for all eigenpairs at each SCF iteration after the first few, 
iterative diagonalization methods such as Davidson~\cite{davi:75} and steepest descent~\cite{long:80,sing:89} can be much more efficient than 
direct methods, 
especially as problem sizes increase and memory constraints become a 
significant 
concern. However, while iterative solution methods make much larger computations possible, diagonalization remains the key bottleneck in large-scale \emph{ab initio} calculations. 
Due to the nonorthogonal basis sets employed in electronic structure methods
such as APW+lo, AO, and PUFE,
the resulting numerical eigenvalue problems 
can be ill-conditioned. In particular,  
$H^{(i_s)}$ and $S$ coefficient matrices
can be ill-conditioned and share a large common near-null
subspace. Furthermore, there is in general no clear gap between the 
eigenvalues that are sought (i.e., occupied states) and the rest. 
Moreover, the ill-conditioning and difficulty of 
iterative diagonalization become especially pronounced as bases become
saturated with orbital functions with long tails in order to 
attain high accuracy.

\bigskip 

Table~\ref{tab:cond} shows the condition
numbers $\kappa(H^{(1)})$ and $\kappa(S)$ 
of coefficient matrices $H^{(1)}$ and $S$, respectively, 
at the first SCF iteration of PUFE calculations
of metallic CuAl, using HGH pseudopotentials~\cite{hart:98}. 
There are two atoms in the triclinic unit cell, which is subject to
Bloch-periodic boundary conditions~\cite{suku:09}. The Brillouin zone is sampled at the $\Gamma$-point and at $\vm{k} = (0.12,-0.24,0.37)$.
The lattice vectors for the unit cell are:
\begin{align}
\vm{a}_1 &= a(1.00,\;  0.02,\; -0.04),\nonumber \\
\vm{a}_2 &= a(0.06,\;  1.05,\; -0.08),\nonumber \\
\vm{a}_3 &= a(0.10,\; -0.12,\;  1.10),\nonumber 
\end{align}
with lattice parameter $a = 5.7$ Bohr. The Cu and Al atoms are 
located at lattice coordinates $\bsym{\tau}_1=(0.01,0.02,0.03)$ and 
$\bsym{\tau}_2=(0.51,0.47,0.55)$,
respectively. Total energy calculations with PUFE are carried out on a
uniform $n_0 \times n_0 \times n_0$ cubic-order finite element mesh,
$r_e$ is the enrichment support radius,  
and $n_{\rm dof}$ is the resulting dimension of the GHEP \eqref{ghep}.

\begin{table}
\begin{center} 
\caption{Condition numbers of $H^{(1)}$ and $S$ matrices 
in PUFE calculations of CuAl
as a function of enrichment support radius.
}\label{tab:cond}
\vspace*{0.1in}
\begin{tabular}{ccccc} \hline
$n_0$ & $r_e$ & $n_{\rm dof}$ & $\kappa(H^{(1)})$ & $\kappa(S)$ \\\hline\hline
6 & 0.0 &     1512 & 7.4e02  & 3.0e03  \\\hline
6 & 1.0 &    1532 & 6.5e07  & 3.5e08  \\\hline
6 & 2.0 &   1685 & 3.3e08  & 3.3e09  \\\hline
6 & 3.0 &   2112 & 5.6e09  & 6.2e10  \\\hline
6 & 4.0 &  2518 & 3.0e10  & 4.5e11  \\\hline
\end{tabular}
\end{center} 
\end{table}

\begin{figure} 
\centering
\includegraphics[width=1.0\textwidth]{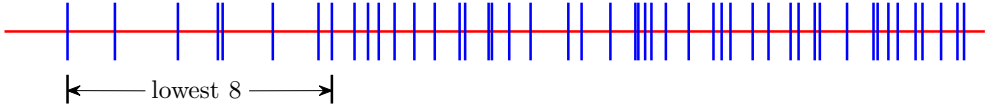}
\vspace*{-0.4in}
\caption{Partial distribution of energy eigenvalues for 
CuAl, showing the clustering and proximity of the lowest 8 computed values to the remainder of the spectrum.} 
\label{fig:ill-cond} 
\end{figure}

In Table~\ref{tab:cond}, the classical FE method corresponds to the case
of no orbital enrichment, i.e., $r_e=0$~\cite{pask:05}. In this case, 
both matrices $H^{(1)}$ and $S$ are well-conditioned.
However, once $r_e > 0$ and orbital enrichments are added, the
condition numbers of $H^{(1)}$ and $S$ increase sharply.
In addition, we observe that $H^{(1)}$ and $S$ share a large 
common near-null subspace. 
For example, 
when $n_0=6$ and $r_e=1.0$,
there is a subspace of dimension $n_e = 20$ 
spanned by the columns of an orthogonal matrix 
$V$ with $\|V\|=1$ such that 
$\|H^{(1)} V \| = \|S V \| = O(10^{-4})$. 
Furthermore, some eigenvalues are clustered and there is no obvious gap between the
eigenvalues of interest and the rest. Figure~\ref{fig:ill-cond} shows the lowest 8 (3\% of the eigenvalues of $H^{(1)}$ and $S$) of interest and higher states in the vicinity.

Ill-conditioned generalized eigenvalue 
problems in quantum mechanical calculations with large nonorthogonal
basis sets have been studied for decades, since the introduction of such bases,
see for example~\cite{lowd:67,jung:92}.
The challenges of solving ill-conditioned problems
arising from the partition-of-unity FE method is an active research area, 
see for example~\cite{stro:06,gers:12}. 
In the next section, we propose a hybrid preconditioning technique
for the rapid iterative diagonalization 
of ill-conditioned GHEPs \eqref{ghep}, as occur in orbital based \tit{ab initio} methods such as APW+lo, AO, and PUFE.

\section{Hybrid preconditioning and LABPSD}\label{sec:labpsd} 

In this section, we consider the rapid iterative diagonalization of the GHEP \eqref{ghep}. 
Specifically, we start with the approximates 
$\{\widehat{E}_0, \widehat{\Psi}_0\}$ 
of the lowest $m$ eigenpairs of \eqref{ghep} from the previous 
SCF iteration. The objective at the $i_s$-th iteration 
is to compute improved approximate eigenpairs  
$\{\widehat{E}_{l}, \widehat{\Psi}_{l}\}$ satisfying \eqref{eq:resL}. 

The block preconditioned steepest descent (BPSD) method \cite{knya:03b}, 
also known as a simultaneous Rayleigh quotient minimization 
method~\cite{long:80}, proceeds as follows. Assume 
$\{\widehat{E}_{\ell-1}, \widehat{\Psi}_{\ell-1}\}$ 
are obtained from $(\ell-1)$-st BPSD iteration with
the residuals
\[
R =H\widehat{\Psi}_{\ell-1}-S\widehat\Psi_{\ell-1}\widehat{E}_{\ell-1}, 
\]
where for simplicity, the superscript 
$i_s$ of $H^{(i_s)}$ is dropped here and in the remainder
of this section.  For the $\ell$-th approximate eigenpairs, 
we first compute search space vectors: 
\[ 
p_i =-{K}_i r_i \quad \mbox{for $i = 1, 2, \ldots, m$,} 
\] 
where $r_i$ is the $i$th column of $R$, and $K_i$ is 
the corresponding preconditioner. 
$p_i$ is also called a \tit{preconditioned residual}. 
Let $Z=[\widehat{\Psi}_{\ell-1} \; P]$ with 
$P = [p_1, p_2, \ldots, p_m]$,  
then the $\ell$-th approximate eigenpairs  
$\{\widehat{E}_{\ell}, \widehat{\Psi}_{\ell}\}$ 
are obtained via the Rayleigh-Ritz   
procedure with the projection subspace matrix $Z$, i.e., 
$\widehat{E}_{\ell} = \Gamma$, 
$\widehat{\Psi}_{\ell}=ZW$, and 
$\{\Gamma,W\}$ are the lowest $m$ eigenpairs of the reduced 
matrix pair $ ({H}_{\rm R}, {S}_{\rm R}) = (Z^H H Z, Z^HSZ)$.   

The convergence of the BPSD method depends critically 
on the preconditioners $K_i$. 
As we have discussed in Section~\ref{sec:scf}, 
due to the nonorthogonal basis sets employed
in electronic structure methods such as APW+lo, AO, and PUFE, 
the GHEP \eqref{ghep} can be ill-conditioned. 
It is well known that 
the presence of large off-diagonal entries in $H$ and $S$ from
local orbital components of such bases render standard preconditioning techniques 
based on the diagonal of $H-\sigma S$ no longer effective~\cite{rays:08,blah:10}.

In the recent work of Blaha {\em et al}~\cite{blah:10} 
on iterative diagonalization in the context of the APW+lo method, 
the following preconditioner is proposed:    
\begin{equation}  \label{globalprecond} 
K_i = (H - \bar{\varepsilon} S)^{-1} 
\equiv  K_{\bar{\varepsilon}} \quad \mbox{for all $i$}, 
\end{equation}
where $\bar{\varepsilon}$ 
is a parameter chosen close to the eigenvalues of interest, 
and the matrices $H$ and $S$ 
are chosen from a fixed (usually the first) SCF iteration
and do not change in the entire SCF procedure.
We call $K_{\bar{\varepsilon}}$ a {\em global preconditioner}. 
Such a global preconditioner has been proposed 
in the context of FE~\cite{ande:05} and 
planewave~\cite{sawa:99} based methods as well.

To apply the global preconditioner \eqref{globalprecond}, 
in \cite{blah:10}, a dense LDL$^{\rm T}$ factorization of 
$K_{\bar{\varepsilon}}$ is first computed
and stored on disk. 
During the entire SCF procedure, the factorization is read in
to perform the required matrix-vector multiplications. 
In \cite{ande:05}, in the context of an FE basis, the search space vectors 
$\{p_i\}$
are computed approximately by an iterative linear solver.
Unfortunately, as we show 
in Section~\ref{sec:num},
such a preconditioner leads to 
stagnation in the context of less well-conditioned PUFE matrices. 

In \cite{wood:85,sing:89,pask:05a,ande:05,rays:08},
the following preconditioners are proposed
to individually target eigenpairs of interest:
\begin{equation} \label{localprecond} 
K_i  = (H-\widehat{\varepsilon}_i S)^{-1} \equiv 
K_{\widehat\varepsilon_i} 
\quad \mbox{for $i = 1, 2, \ldots, m$}, 
\end{equation}
where $\widehat{\varepsilon}_i$ are 
Ritz values from the previous BPSD iteration, 
i.e., diagonal elements of $\widehat{E}_{\ell-1}$. 
The basic motivation can be understood as follows (e.g., \cite{wood:85,sing:89,davi:89}). 
Given current approximation $\{\psih_i,\epsh_i\}$ to eigenpair $\{\psi_i,\eps_i\}$, we seek correction $p_i$ such that $\psih_i+p_i$ is exact, i.e., 
\begin{equation}
\label{eq:exactpi}
(H - \eps_i S)(\psih_i + p_i) = 0.
\end{equation}
While the exact eigenvalue $\eps_i$ is unknown, the Rayleigh quotient
\begin{equation}
\epsh_i=\frac{\psih_i^H H \psih_i}{\psih_i^H S \psih_i}
\end{equation}
provides an excellent approximation, with an error that is second order in the error of $\psih_i$. Replacing $\eps_i$ with $\epsh_i$ in \eqref{eq:exactpi} then gives
\begin{equation}
(H - \epsh_i S)(\psih_i + p_i) = r_i + (H - \epsh_i S) p_i = 0
\end{equation}
or
\begin{equation}
\label{eq:approxpi}
p_i = -(H - \epsh_i S)^{-1} r_i,
\end{equation}
as in \eqref{localprecond}. Note, however, that as $\epsh_i$ approaches $\eps_i$, the matrix $H - \epsh_i S$ becomes singular and hence the inverse exists only in the subspace orthogonal to $\psi_i$ and any vectors degenerate with it~\cite{sing:89,davi:89,slei:00}. Furthermore, for $\epsh_i \ne \eps_i$, the inverse exists and returns the correction $p_i = -\psih_i$, providing no correction to the direction of $\psih_i$ whatsoever. In practice, since the inverse is computed only approximately, neither of these issues is a particular 
concern; 
however, they can affect convergence at higher accuracies~\cite{slei:00}. 
In the present case, we 
solve the equation 
\begin{equation} \label{eq:idealp} 
( H-\widehat\varepsilon_i S ) p_i = -r_i,   
\end{equation}  
\tit{inexactly}, i.e., find $\widehat{p}_i$ satisfying 
\begin{equation} \label{eq:approxp} 
\|(H-\widehat\varepsilon_i S) \widehat{p}_i + r_i \|
\leq \eta \|r_i\|, 
\end{equation}  
where $\eta$ is a prescribed tolerance.

An asymptotic analysis of superlinear convergence of
the preconditioner $K_{\widehat\varepsilon_1}$ for computing the smallest 
eigenpair has been studied in~\cite{samo:58,ovtc:06}. 
This convergence analysis is extended for the case of 
multiple eigenpairs in our recent work~\cite{cai:12b}. 
Since the preconditioners
$\{ K_{\widehat{\varepsilon}_i} \}^m_{i=1}$ 
accelerate the convergence of individual eigenpairs $\{\psih_i,\epsh_i\}$, we refer to them as {\em locally accelerated} preconditioners. 

It is a computational challenge to 
apply the locally accelerated preconditioners at each BPSD iteration
in a cost-effective way. In \cite{rays:08}, it is suggested to 
first compute the full spectral decomposition of 
the matrix pair $(H,S)$ at some SCF iteration. 
However, the spectral decomposition is prohibitively
expensive for large-scale systems.
In \cite{pask:05a,ande:05},
the conjugate-gradient method is used for solving \eqref{eq:idealp}. 
This allows for very larger-scale calculations.
However, the CG method (or MINRES for indefinite systems) 
suffers slow convergence and stagnation due to the 
ill-conditioning of the coefficient matrices, in the PUFE context in particular.

To overcome the slow convergence of higher eigenpairs using the 
global preconditioner and high computational cost and stagnation 
of the locally accelerated preconditioners, we propose the 
following \tit{hybrid preconditioning} scheme: 
\begin{enumerate} 
\item In the initial few SCF iterations, 
apply only the global preconditioner $K_{\bar{\varepsilon}}$ 
to compute all search space vectors $P = [p_1, p_2, \ldots, p_m]$, 
i.e.,
\[ 
P = -K_{\bar{\varepsilon}}R 
= -(H - \bar{\varepsilon} S)^{-1} R. 
\] 
\item If the $i$-th approximate eigenvalue $\widehat{\varepsilon}_i$ 
is {\em localized}, apply the locally accelerated preconditioner  
$K_{\widehat{\varepsilon}_i}$ in two stages:
\begin{enumerate} 
\item Compute an initial search space vector $\widehat{p}^{(0)}_i$
by applying the global preconditioner $K_{\bar{\varepsilon}}$: 
\[
\widehat{p}^{(0)}_i 
= -K_{\bar{\varepsilon}}r_i 
= -(H - \bar{\varepsilon} S)^{-1} r_i. 
\]
\item Iteratively refine $\widehat{p}^{(0)}_i$ to find the search space vector
$\widehat p_i$ satisfying~\eqref{eq:approxp}.  
\end{enumerate} 
\end{enumerate} 
This two-stage application of locally accelerated preconditioners
$K_{\widehat\varepsilon_i}$ addresses the issue of slow convergence 
of iterative methods for computing $\widehat p_i$.
Using a good initial approximation $\widehat{p}^{(0)}_i$,  
the iterative refinement is expected to converge in just a 
few iterations, typically 2 to 5. The pre-application of the global preconditioner is
efficient since the factorization of the global preconditioner
is already available from the initial SCF iteration.
As shown in 
Section~\ref{sec:num}, 
the proposed hybrid preconditioning scheme amortizes the cost of the global preconditioner 
and significantly reduces the cost of the more aggressive locally accelerated preconditioners, 
yielding a {\em cost-effective} preconditioning scheme for 
the iterative diagonalization of ill-conditioned GHEPs.

We shall refer to the combined algorithm, BPSD with above hybrid preconditioning, as the 
{\em Locally Accelerated Block Preconditioned Steepest Descent} (LABPSD) method. 
An outline of the method is as follows:
\begin{quote} 
\begin{enumerate}
\item Input initial approximate eigenpairs 
$\{E,\Psi\}$, where $\Psi \in\mathbb{C}^{n\times (m+m_0)}$

\item Compute ${\rm tol}_0 = \mbox{Res}_{\max}[{E}, {\Psi}]$ 

\item Compute matrix-vector products $\Psi_H=H\Psi$ and $\Psi_S=S\Psi$ 

\item Compute residual vectors
      $R={\Psi_H}_{(:,1:m)}-{\Psi_S}_{(:,1:m)}E_{(1:m,1:m)}$

\item Test for convergence to tolerance $\tau^{(i_s)}_{\rm eig}$. If converged, exit

\item Set up search subspace 
      $Z = [\Psi \,\, P]$ with preconditioned residual vectors $P$ 
      computed as follows: 
  
      (a) Apply global preconditioner: $P = - K^{(i_0)}_{\bar\varepsilon} R$
 
      (b) If $\varepsilon_i = E_{(i,i)}$ is localized for some $i$ and 
$1\leq i \leq m$, 
refine $p_i = P_{(:,i)}$ with locally accelerated preconditioner, 
i.e., compute correction vector $\delta p_i$ by solving 
refinement equation   
\[ 
(H - \varepsilon_i S)\delta p_i  = -\delta{r}_i
\] 
{\em inexactly}, where $\delta {r}_i = (H- \bar\varepsilon S)p_i  + r_i$. 
Set $ P_{(:,i)} := p_i + \delta p_i$  

\item Perform matrix-vector products $P_H=H P$ and $P_S=SP$

\item Set up coefficient matrices of reduced GHEP 
\[ 
H_{\rm R} = [\Psi \,\, P]^H[\Psi_H \,\, P_H]
\quad \mbox{and} \quad
S_{\rm R} = [\Psi \,\, P]^H[\Psi_S \,\, P_S]
\] 

\item Compute lowest $m+m_0$ eigenpairs 
$\{W,E\}$ of $({H}_{\rm R},{S}_{\rm R})$: 
\[ 
{H}_{\rm R} W = {S}_{\rm R} W E
\] 

\item Compute new approximate eigenvectors $\Psi :=[\Psi\, P]W$ 

\item Update $\Psi_H :=[\Psi_H \, P_H]W$ and $\Psi_S :=[\Psi_S\, P_S]W$

\item Go to step~4.
\end{enumerate}
\end{quote} 

A few remarks are in order. 
\begin{enumerate} 
\item 
The initial approximations
$\Psi$ are eigenvectors $\Psi^{(i_s-1)}$ 
from the previous SCF iteration, i.e., $\Psi = \Psi^{(i_s-1)}$.
Having the extra $m_0$ vectors is important. It can accelerate convergence 
substantially when there are multiple (degenerate) or clustered eigenvalues at or near the $m$-th.
In practical calculations (with multiplicities limited by symmetries in the underlying physical problem), a small $m_0$ is generally sufficient, for example $m_0 = m/10$. 
The larger the $m_0$, the faster the convergence, but also the more matrix-vector products required. 
Similar findings pertain for other solvers in the electronic structure context as well, see for example~\cite{kress:96,rays:08,blah:10}.

\item The LABPSD iteration is considered to be converged if 
$\mbox{Res}_{\max}[{E}_{(1:m,1:m)}, {\Psi}_{(:,1:m)}] \leq \tau^{(i_s)}_{\rm eig}$.

\item Line 6 is only executed for 
residual vectors corresponding to 
unconverged eigenpairs. 
The implementation details are presented in Section~\ref{sec:impl}.

\item The $i$-th approximate eigenpair
$\{\varepsilon_i, \psi_i\} = \{E_{(i,i)}, {\Psi}_{(:,i)}\}$
is deemed ``localized'' if the following conditions
are satisfied: 
\[ 
\mbox{Res}[\varepsilon_i, \psi_i] \leq \tau_1 
\quad \mbox{and} \quad 
|\varepsilon_i - \varepsilon^{\ell-1}_i| \leq  
\tau_2 |\varepsilon^{\ell-1}_i|, 
\]
where $\varepsilon^{\ell-1}_i$ is the $i$-th approximate
eigenvalue from the previous ($\ell-1$) BPSD iteration. 
Both $\tau_1$ and $\tau_2$ are parameters.
In our numerical tests, we set $\tau_1=\tau_2=0.1$. 
The above localization condition thus provides an indication that the $i$-th approximate eigenvalue $\varepsilon_i$ has settled down sufficiently with respect to BPSD iterations $\ell$ to be used as a shift for preconditioning.

\item 
By storing the block vectors $\Psi_H$, $\Psi_S$, $P_H$ and $P_S$, 
the matrices $H$ and $S$ are accessed only once per BPSD iteration, 
other than in preconditioning step 6.  

\item The reduced dense GHEP $({H}_{\rm R},{S}_{\rm R})$ 
can be solved by standard routines such as LAPACK {\tt ZHEGVX}.
\end{enumerate} 

\section{Implementation details} \label{sec:impl}
In this section, we discuss implementation details of
the hybrid preconditioning scheme in step 6 of the LABPSD method. 

First, we consider the global preconditioning step 6(a).
As discussed in Section~\ref{sec:labpsd}, the global preconditioner
$K_{\bar\varepsilon}$ is fixed throughout the SCF iterations. 
Typically, the coefficient matrices 
$H^{(1)}$ and $S$ in the first SCF iteration 
are sufficient to construct an effective $K_{\bar\varepsilon}$, i.e., $i_0=1$ in line 6(a) of LABPSD. 
Therefore, let us consider how to exploit the structure 
of $H^{(1)}$ and $S$ to efficiently compute 
\begin{equation} \label{eq:k1p} 
P =-K^{(1)}_{\bar\varepsilon} R 
  = -\left(H^{(1)} - \bar\varepsilon S \right)^{-1}R. 
\end{equation} 
From the definition \eqref{ghep} of $H^{(1)}$, 
the global preconditioner $K^{(1)}_{\bar\varepsilon}$ is
the inverse of a Hermitian matrix plus low-rank update:   
\begin{equation}  \label{eq:k1} 
K^{(1)}_{\bar\varepsilon} 
=\left( H^{(\loc,1)}- \bar\varepsilon S + H^{(\nl)}\right)^{-1},
\end{equation} 
where $H^{(\loc,1)}-\bar\varepsilon S$ is Hermitian 
and $H^{(\nl)}$ has the rank-revealing decomposition
\begin{equation}  \label{eq:fgf} 
H^{(\nl)}=FGF^H,
\end{equation}
where $F$ is $n$-by-$k$ and $G$ is $k$-by-$k$ Hermitian. 
The rank $k$ is the number of projectors in the pseudopotential formulation, 
typically $k \ll n$. 
For localized bases such as PUFE, $H^{(\loc,1)}$ and $S$ are sparse.\footnote{In PUFE, $H^{(\loc,1)}$ and $S$ share the same sparsity pattern.} 

To compute $P$, we first compute the following factorization
of the matrix $H^{(\loc,\; 1)} - \bar\varepsilon S$:
\begin{equation}  \label{eq:ldl} 
\Pi^\top (H^{(\loc,\; 1)} - \bar\varepsilon S) \Pi = LDL^H,  
\end{equation}
where $\Pi$ is a permutation matrix, $L$ is a unit lower triangular matrix, 
and $D$ is a block diagonal matrix with only 1-by-1 and 2-by-2 blocks
on the diagonal. 
Algorithms for the factorization \eqref{eq:ldl} 
are well-established, see for example~\cite{sche:04,sche:06,davi:05}.
Since the global preconditioner is unchanged during the SCF iterations,
the factorization \eqref{eq:ldl} is computed just once and
used throughout the SCF process.
This is along the lines of the global preconditioning scheme 
suggested in \cite{blah:10}. However, in the context of a localized basis and sparse matrices, such as PUFE, we use a sparse factorization rather than dense one as in \cite{blah:10}. 

With the low-rank representation \eqref{eq:fgf} and factorization \eqref{eq:ldl},
we can compute the global-preconditioned search space vectors $P$ 
using the Sherman-Morrison-Woodbury (SMW) formula \cite{govl:96} as follows:  
\begin{quote}
\begin{enumerate}
\item Compute $\widehat{F}= (H^{(\loc,\; 1)} - \bar\varepsilon S)^{-1}F$
using the factorization~\eqref{eq:ldl} 

\item $F:=FG$ 

\item $T=I+F^H\widehat{F}$

\item $F:= F T^{-H}$ 

\item Compute $P = -(H^{(\loc,\; 1)} - \bar\varepsilon S)^{-1}R$
using the factorization~\eqref{eq:ldl} 

\item $P :=P -\widehat{F}F^H P$
\end{enumerate}
\end{quote}
Here we have arranged the order of computations such that 
the first four steps are executed just once. 
By storing $F$ and $\widehat F$, 
$P$ can be computed using only the last two steps. 

Turning now to the locally accelerated preconditioning step 6(b),
the iterative refinement of initial approximate $\widehat{p}^{(0)}_i = P_{(:,i)}$ 
computed in step 6(a) can be recast as solving the following linear system:
\begin{equation} \label{eq:indls} 
(H^{(i_s)} - \varepsilon_i S) {p}_i  = -r_i,
\end{equation}
with starting vector $\widehat{p}^{(0)}_i$.
Since $H^{(i_s)}-\varepsilon_i S$ is Hermitian and indefinite, 
MINRES \cite{paig:75, vors:03} is a natural choice. 
Although the coefficient matrix $H^{(i_s)}-\varepsilon_i S$ 
of \eqref{eq:indls} can become highly ill-conditioned, as we show below,
we observe that it takes just 2 to 5 iterations 
for MINRES to converge to the desired tolerance starting from the 
pre-processed vectors $\widehat{p}^{(0)}_i$ from the global preconditioner.

\section{Results} \label{sec:num} 

In this section, we provide numerical results to 
demonstrate the efficiency of the LABPSD algorithm 
for the rapid iterative diagonalization 
of ill-conditioned generalized eigenvalue problems 
produced by the PUFE electronic structure method~\cite{suku:09,pask:11,pask:11a}, which employs a strictly local nonorthogonal basis combining atomic orbitals for efficiency and finite elements for generality and systematic improvability.

We have conducted extensive tests of the LABPSD method
in PUFE calculations of a variety of materials systems. 
Here we show results for two systems representative of opposite extremes: CuAl with a soft, shallow pseudopotential and clustered or degenerate eigenvalues, and CeAl with a notably hard and deep pseudopotential and nondegenerate spectrum.

\paragraph{CuAl} 
Our first test case is a high-symmetry, cubic CuAl metallic system, with $\Gamma$-point Brillouin zone sampling to maximize degeneracies in the spectrum.
The unit cell is body-centered cubic with 
lattice parameter $a = 5.8$ Bohr and atomic positions 
$\bsym{\tau}_1=(0.0,0.0,0.0)$ (Cu) and 
$\bsym{\tau}_2=(0.5,0.5,0.5)$ (Al), in lattice coordinates.  
The Brillouin zone is sampled at the $\Gamma$-point to maximize degeneracies in the spectrum, including degeneracy at the Fermi level, thus providing a stringent test of the eigensolver's ability to extract clustered/degenerate eigenpairs.
The resulting spectrum has a triple-degeneracy 
(eigenvalues of $0.36047$ Hartree) and
a double-degeneracy (eigenvalues of $0.37553$ Hartree), which is also the highest
occupied state with Fermi-Dirac occupation and $k_B T = 0.01$ a.u.

\begin{figure}
\centering
\mbox{
\subfigure[]{\epsfig{file=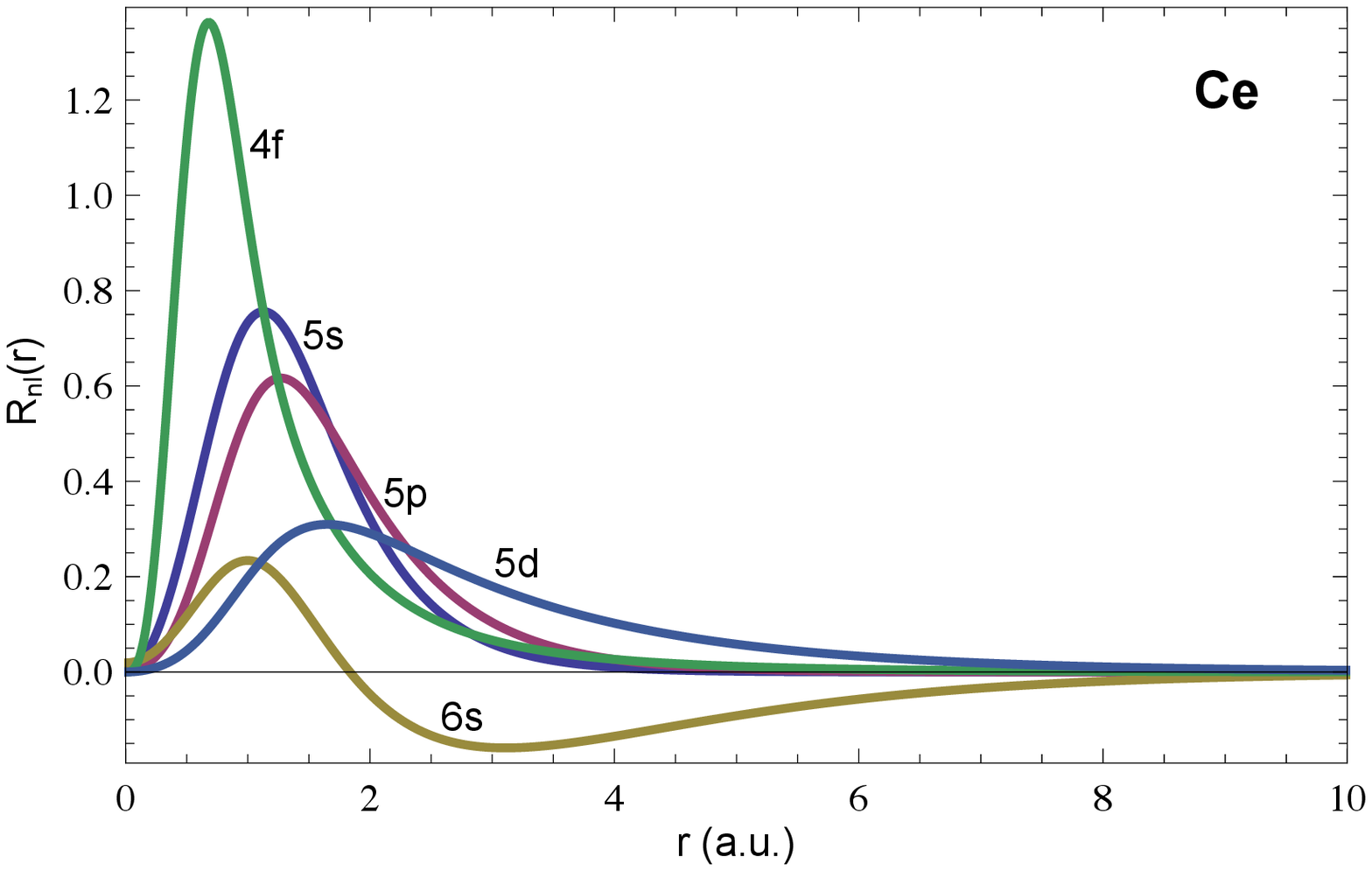,width=0.48\textwidth}}
\subfigure[]{\epsfig{file=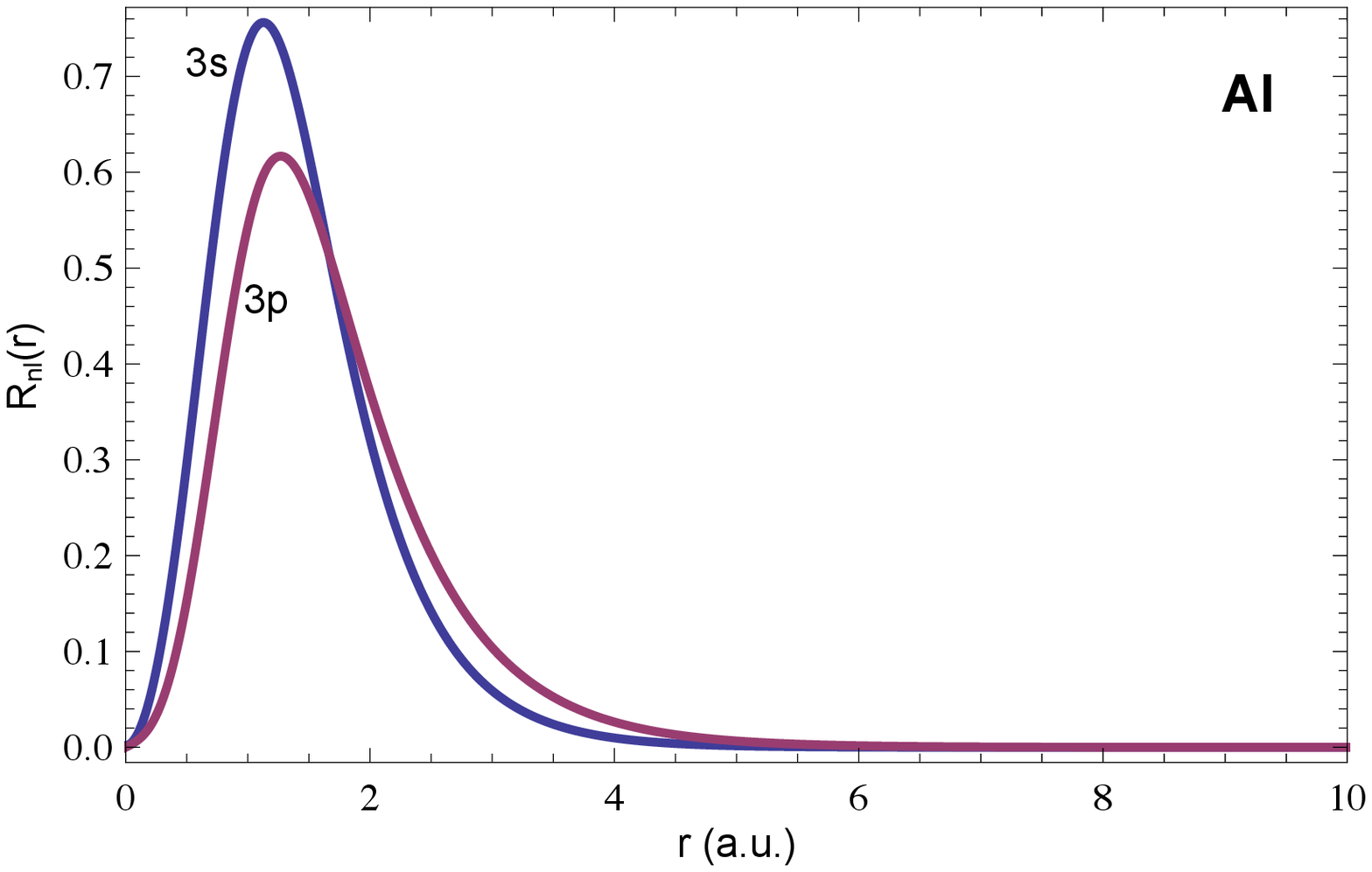,width=0.48\textwidth}}
}
\caption{PUFE orbital enrichment functions for CeAl. Radial parts for
(a) Ce and (b) Al.}
\label{fig:ceal}
\end{figure}

\paragraph{CeAl} 
As a test of the solver's ability to handle general, nondegenerate spectra, with low-lying  eigenvalues and thus broader overall spectrum, we consider next the case of metallic, triclinic CeAl. This is a particularly challenging system due to the following properties:
(a) The potentials of the atoms are deep, producing strongly
localized solutions, with low-lying eigenvalues, that require larger basis sets to resolve.
(b) The atoms are heavy, with many electrons in valence,
requiring many eigenfunctions to be computed.
(c) Because the system contains Ce, it requires 17 orbital enrichment functions
per atom (in contrast to Cu for example, which requires only 1),
which increases basis size
substantially for PUFE. The radial parts of the orbital enrichment functions 
(pseudoatomic wavefunctions) for
Ce and Al are shown in Figure~\ref{fig:ceal}. 
(d) The lattice is triclinic with atoms displaced from ideal
positions. This provides a completely general problem, with
no special symmetries to exploit and general, nondegenerate spectrum.
(e) We do not assume a band gap, but rather
solve the general metallic problem with Fermi-Dirac occupation and $k_B T = 0.01$ a.u.

The triclinic unit cell for CeAl has lattice vectors
\begin{align}
\vm{a}_1 &= a(1.00,\;  0.02,\; -0.04),\nonumber \\
\vm{a}_2 &= a(0.01,\;  0.98,\;  0.03),\nonumber \\
\vm{a}_3 &= a(0.03,\; -0.06,\;  1.09),\nonumber 
\end{align}
with lattice parameter $a = 5.75$ Bohr. Atomic lattice coordinates
are $\bsym{\tau}_1=(0.01,0.02,0.03)$ (Ce) and 
$\bsym{\tau}_2=(0.51,0.47,0.55)$ (Al). 
The Brillouin zone is sampled at the
$\Gamma$-point and at $\vm{k} = (0.12,-0.24,0.37)$.
Therefore, there are two independent 
sequences of GHEPs in the SCF procedure. 

\bigskip

For all simulations, the SCF procedure
is terminated at the $i_s$-th iteration if
the relative change of input and output effective potentials
satisfies
\begin{equation} \label{eq:scfstop}
v_{\rm dif}^{(i_s)}=
\frac{\|V^{\rm out}_\eff-V^{\rm in}_\eff\|}
{\|V^{\rm in}_\eff\|}  \leq \tau_{\rm scf}
\end{equation}
for a prescribed tolerance $\tau_{\rm scf}$.
As reference, total energies are 
also calculated by the {\sc abinit} planewave code~\cite{bott:08} 
with well-converged planewave cutoff. 
By virtue of the orbital functions in the PUFE basis, the dimension of the PUFE GHEP is about a factor of 5 smaller than that of a planewave calculation of the same accuracy.

The dimension of the GHEP \eqref{ghep} is
$n_{\rm dof} = 7 n_0^3 + n_e$,
where $n_0$ is the number of elements the
$x$-, $y$- and $z$-directions (uniform FE mesh) and
$n_e$ is determined by the enrichment support
radius $r_e$ and number of atoms.
The factor of 7 is due to the use of cubic
serendipity brick elements~\cite{suku:09}.
By introducing a shift $\sigma_0$,
$H^{(i_s)} := H^{(i_s)}- \sigma_0 S$ is made Hermitian positive
definite.\footnote{ 
Usually, $\sigma_0$ is selected close to the eigenvalues of interest.
In electronic structure calculations,
a good estimate of the lowest eigenvalue is generally available
so that a shift $\sigma_0$ to make $H^{(i_s)}$ positive definite is readily determined.
}
The PUFE code provides the routines to perform the
matrix-vector multiplications $H^{(\loc)}v$, $H^{(\nl)}v$ and $Sv$
for an arbitrary vector $v$.
Subsequently, the matrix-vector multiplication
$(H^{(i_s)}-\sigma S)v$ is readily computable for
any shift $\sigma$ to facilitate preconditioning.

In addition, 
termination criteria for the SCF, BPSD, and MINRES iterations 
are $\tau_{\rm scf}= 10^{-5}$, 
$\tau^{(i_s)}_{\rm eig}= \frac{1}{10} v^{(i_s)}_{\rm dif}$,
and 
$\eta^{(i_s)} = \eta = 0.25$, respectively. 
The maximum number of outer BPSD and inner MINRES iterations
are set to 20, unless otherwise specified.  
The outermost SCF iterations are repeated until 
convergence of the potential as defined in
\eqref{eq:scfstop} is achieved.  
The global shift $\bar{\varepsilon}$ is chosen 
to be close to the desired eigenvalues of $(H^{(i_s)},S)$. 
In particular, $\bar{\varepsilon} = -0.3$ for CuAl, $\bar{\varepsilon} = -1.0$ for CeAl, which 
are smaller than the estimated smallest eigenvalues 
of $(H^{(i_s)},S)$ for the cases considered here.
As observed in \cite{blah:10}, our numerical experiments also show
$\bar{\varepsilon}$ has little influence 
on the convergence of the BPSD iteration.  

Computations reported in this paper 
were carried out on a two-socket six-core 
Intel Xeon 2.93 GHz processor with 94 GB memory. 
Intel MKL was used for BLAS and LAPACK operations 
in the LABPSD method.  In addition, 
the DSS package of MKL was used
for computing the sparse factorization \eqref{eq:ldl} 
of the global preconditioner. 
DSS is an interface to PARDISO~\cite{sche:04,sche:06} and 
provides subroutines to compute $(H^{(\loc,1)}-\bar\varepsilon S)^{-1}v$ 
for a given vector $v$ after the sparse factorization is computed.

\subsection{SCF convergence} \label{eg1}
We first examine the convergence of the SCF procedure using LABPSD
for the iterative diagonalization of the associated sequence of GHEPs.

\paragraph{CuAl} 
A uniform $12 \times 12 \times 12$ finite-element mesh 
and enrichment support radius $r_e=4$ are employed to provide high accuracy and a strong test of ill-conditioning. 
The dimension of the GHEP \eqref{ghep} is
$n_{\rm dof}=7\times 12^3 + 8130 = 20226$.
The rank of $H^{(\rm nl)}$ is $k=19$.
$m = 10$ eigenpairs are computed in order to accommodate
all electrons in valence and achieve convergence of the effective
potential to the desired accuracy.

\begin{figure}
\centering
\includegraphics[width=0.48\textwidth]{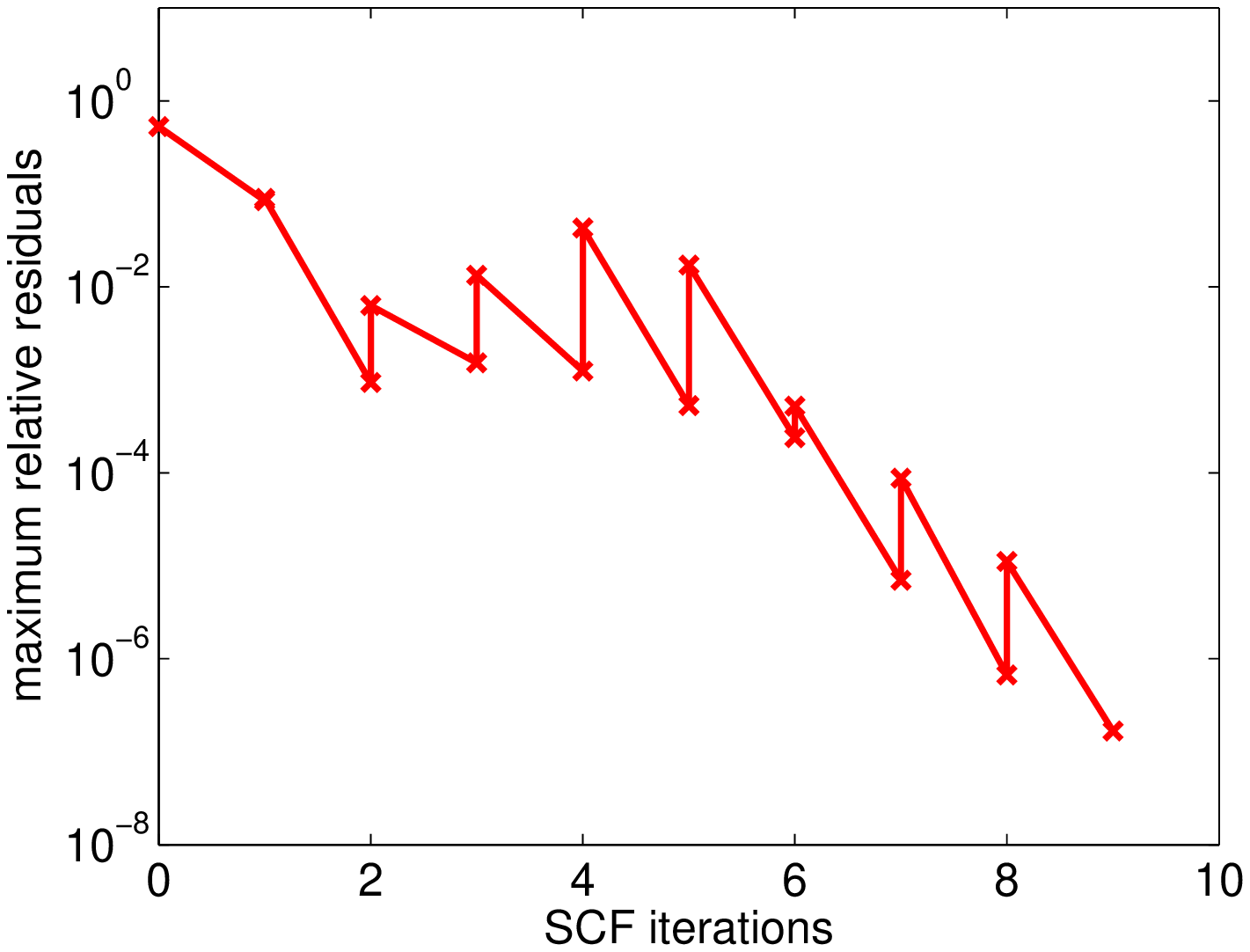}
\includegraphics[width=0.48\textwidth]{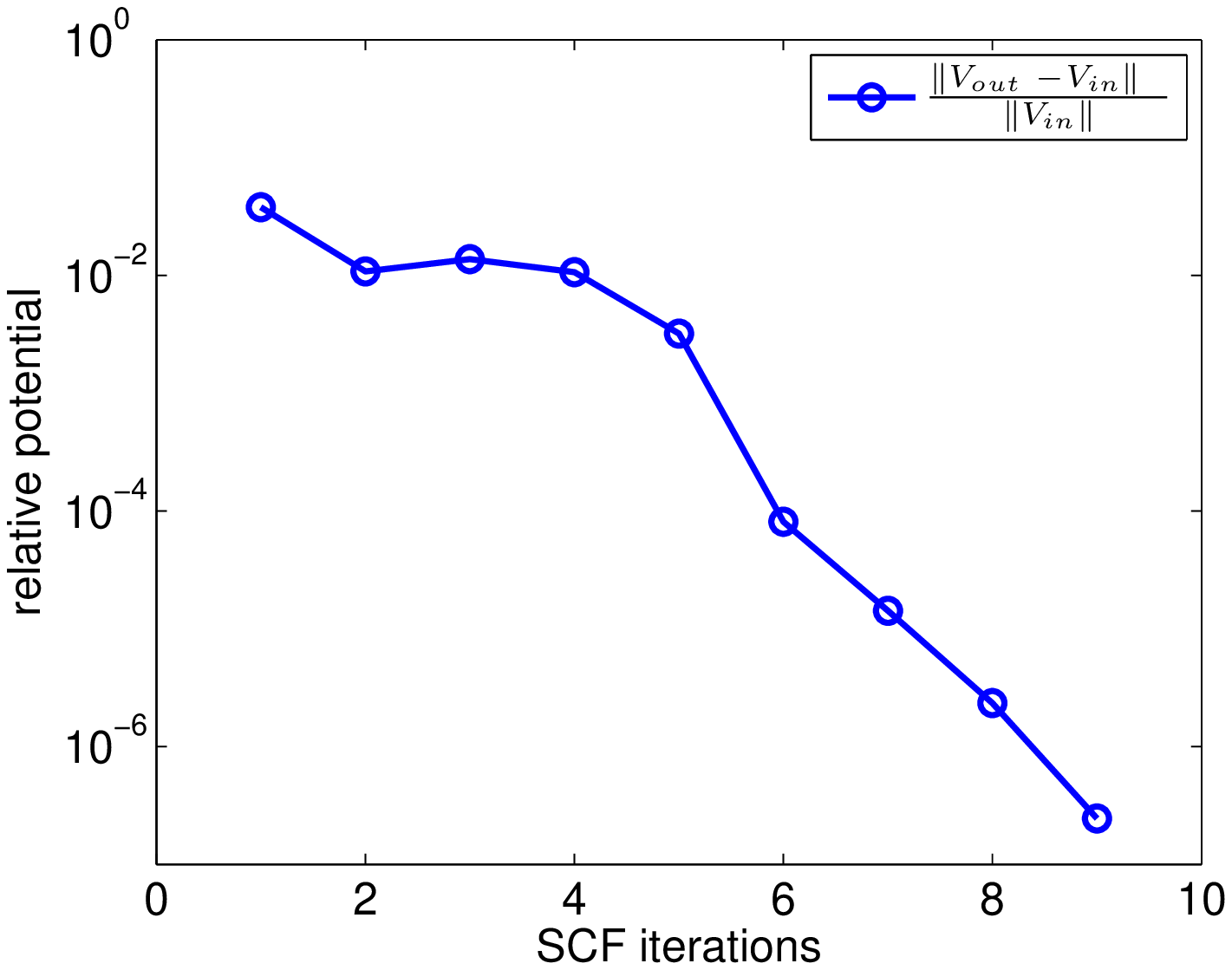}
\caption{CuAl simulation. The maximum relative
backward error of the sequence of 
GHEPs in the solution of NLEP \eqref{ks-nlep} (left), 
and the relative difference
of effective potentials $V_\eff$ (right).} \label{fig:ghepveff1}
\end{figure}

The left plot of Figure~\ref{fig:ghepveff1}
shows the maximum relative
residual errors $\mbox{Res}_{\max}[\widehat{E}, \widehat{\Psi}]$
of the sequence of the GHEPs at the beginning and
end of each SCF iteration, where $m_0 = 10$ for the BPSD iterations.
The right plot of Figure~\ref{fig:ghepveff1}
shows the corresponding difference
$v^{(i_s)}_{\rm dif}$ of input and output effective potentials
(Eq.~\eqref{eq:vdif}).
As can be seen, with LAPBSD as the eigensolver,
the maximum relative residual error of the GHEP steadily drops at the rate
$\tau^{(i_s)}_{\rm eig} = \frac{1}{10} v^{(i_s)}_{\rm dif}$,
along with the input-output potential difference. If the accuracy of the eigensolves at each SCF iteration is further increased, the convergence of the effective potential is not substantially affected.

We note that in the final SCF iteration, the lowest 10 computed eigenvalues are
\begin{center} 
\begin{tabular}{rrrrr}  
 -0.1987515094, & 0.3604669213, & 0.3604669241,& 0.3604669358,& 0.3755287169,\\ 
  0.3755287473, & 0.5721570004, & 0.8464957683,& 0.8464958151,& 0.8464958184,
\end{tabular}
\end{center} 
with triply degenerate value at $\sim 0.3604669$ and doubly degenerate value at $\sim 0.3755287$, as in the reference planewave calculations (deviations from exact degeneracy in the final digits are due to the lower symmetry of the basis than the crystal \cite{PasKleSte99}). As can be seen, the degenerate values pose no particular difficulty for the LABPSD solver.

\begin{figure}
\centering
\includegraphics[width=0.48\textwidth]{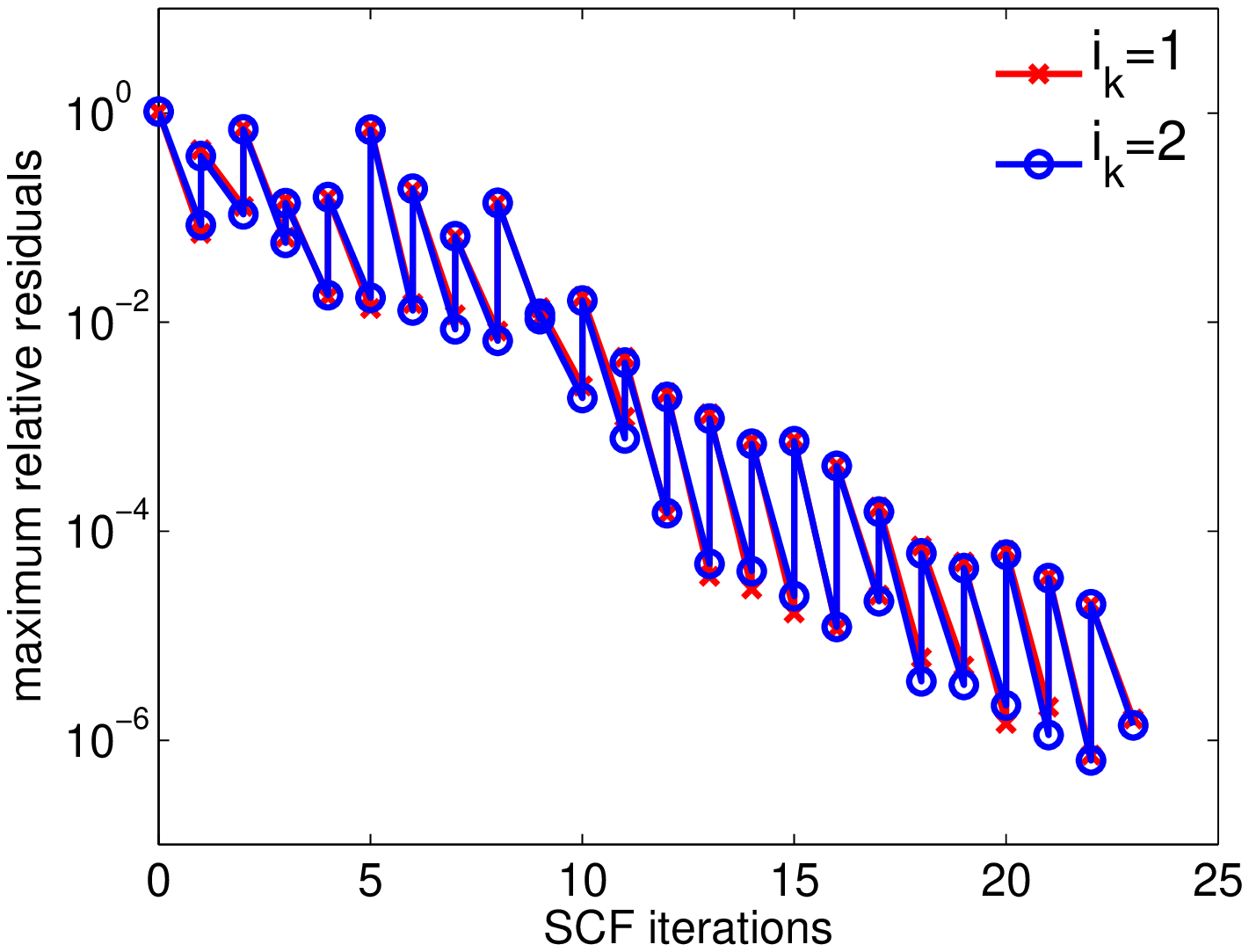}
\includegraphics[width=0.48\textwidth]{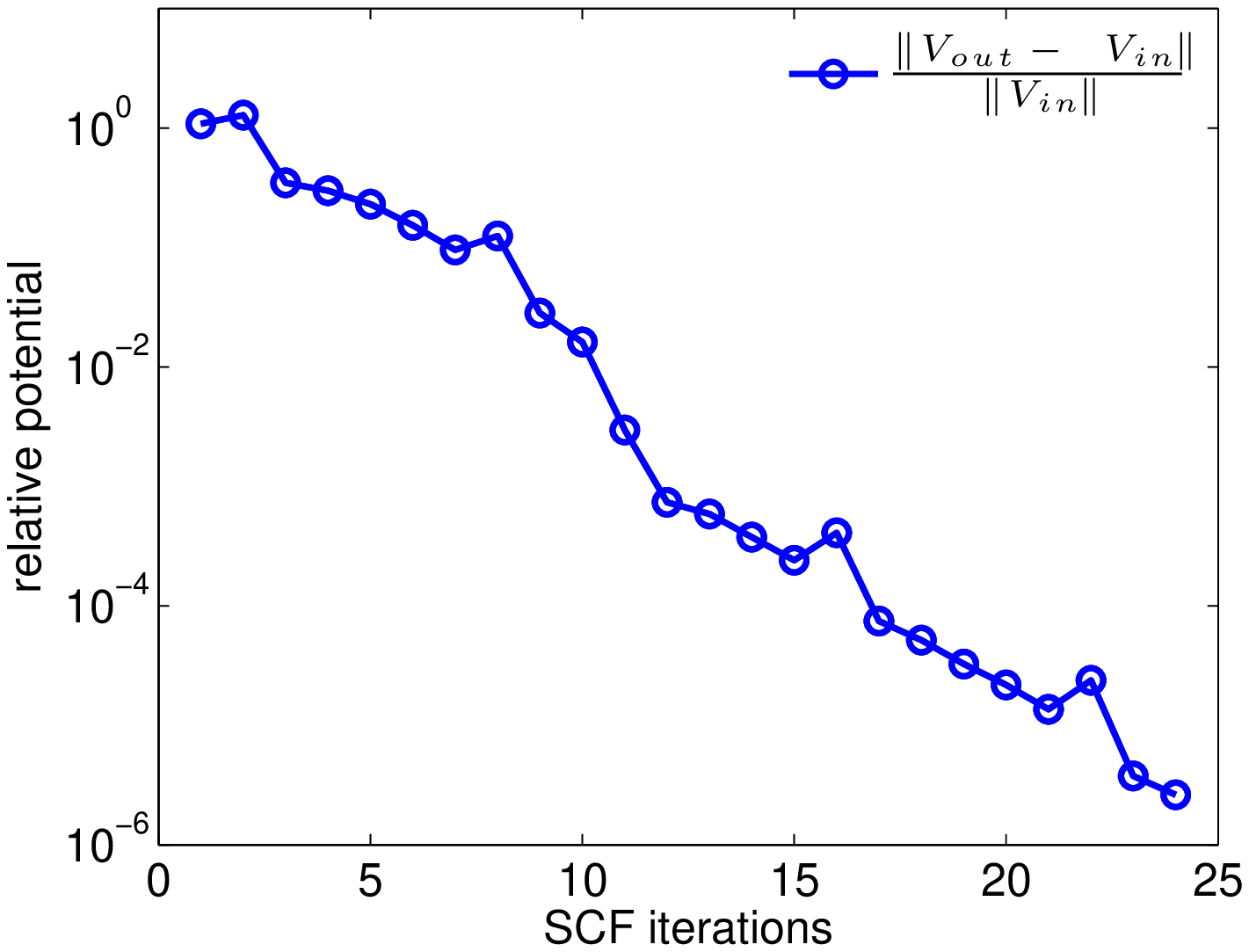}
\caption{CeAl simulation. The maximum relative
backward error of two sequences ($i_k=1$ and $i_k=2$) of
GHEPs in the solution of NLEP \eqref{ks-nlep} (left),
and the relative difference
of effective potentials $V_\eff$ (right).} \label{fig:ghepveff}
\end{figure}

\paragraph{CeAl} For the CeAl system, we consider 
a  $12 \times 12 \times 12$ finite-element mesh with $r_e=2.5$. 
The dimension of the GHEP \eqref{ghep} is $n_{\rm dof}=23795$.
The rank of $H^{(\rm nl)}$ is $k=26$.
In this case, $m = 22$ eigenpairs are computed to accommodate all valence electrons with  specified Fermi-Dirac occupation.
The left plot of Figure~\ref{fig:ghepveff} 
shows the reduction of the maximum relative
residual errors $\mbox{Res}_{\max}[\widehat{E}, \widehat{\Psi}]$
of the sequence of the GHEPs, where $m_0 = 3$. 
The right plot of Figure~\ref{fig:ghepveff} 
shows the corresponding difference
$v^{(i_s)}_{\rm dif}$. 
Again, the maximum relative residual error of the GHEP 
steadily drops at the rate 
$\tau^{(i_s)}_{\rm eig}= \frac{1}{10} v^{(i_s)}_{\rm dif}$,
along with the input-output potential difference.

If the accuracy of the eigensolves at each SCF iteration is further increased, the convergence of the effective potential is not substantially affected.

\subsection{Inner and outer iterations}
\label{sec:iterations}
Now, let us examine the efficiency of 
the LABPSD in terms of the following two quantities: 
\[
{\rm Hp} = \frac{B_s}{S_s \times N_k} 
\quad \mbox{and} \quad 
{\rm La}  = \frac{M_s}{B_s \times m},
\]
where 
$S_s$ is the total number of SCF iterations,  
$B_s$ is the total number of BPSD iterations, 
and $M_s$ is the total number of MINRES iterations. 
$N_k$ is the number of $\vm{k}$-points 
($N_k = 1$ in the CuAl case, $N_k = 2$ in the CeAl case). 
By the above definition, 
${\rm Hp} $ is the average number of outer BPSD iterations per
SCF iteration for each $\vm{k}$-point.  A small ${\rm Hp}$-number indicates the 
efficiency of the hybrid preconditioning technique. 
Similarly, ${\rm La}$ is the average
number of inner MINRES iterations per outer BPSD iteration
for each eigenpair. A small ${\rm La}$-number indicates 
the efficiency of applying the locally accelerated preconditioners 
in the proposed two stages.

\begin{figure}
\centering
\includegraphics[width=0.48\textwidth]{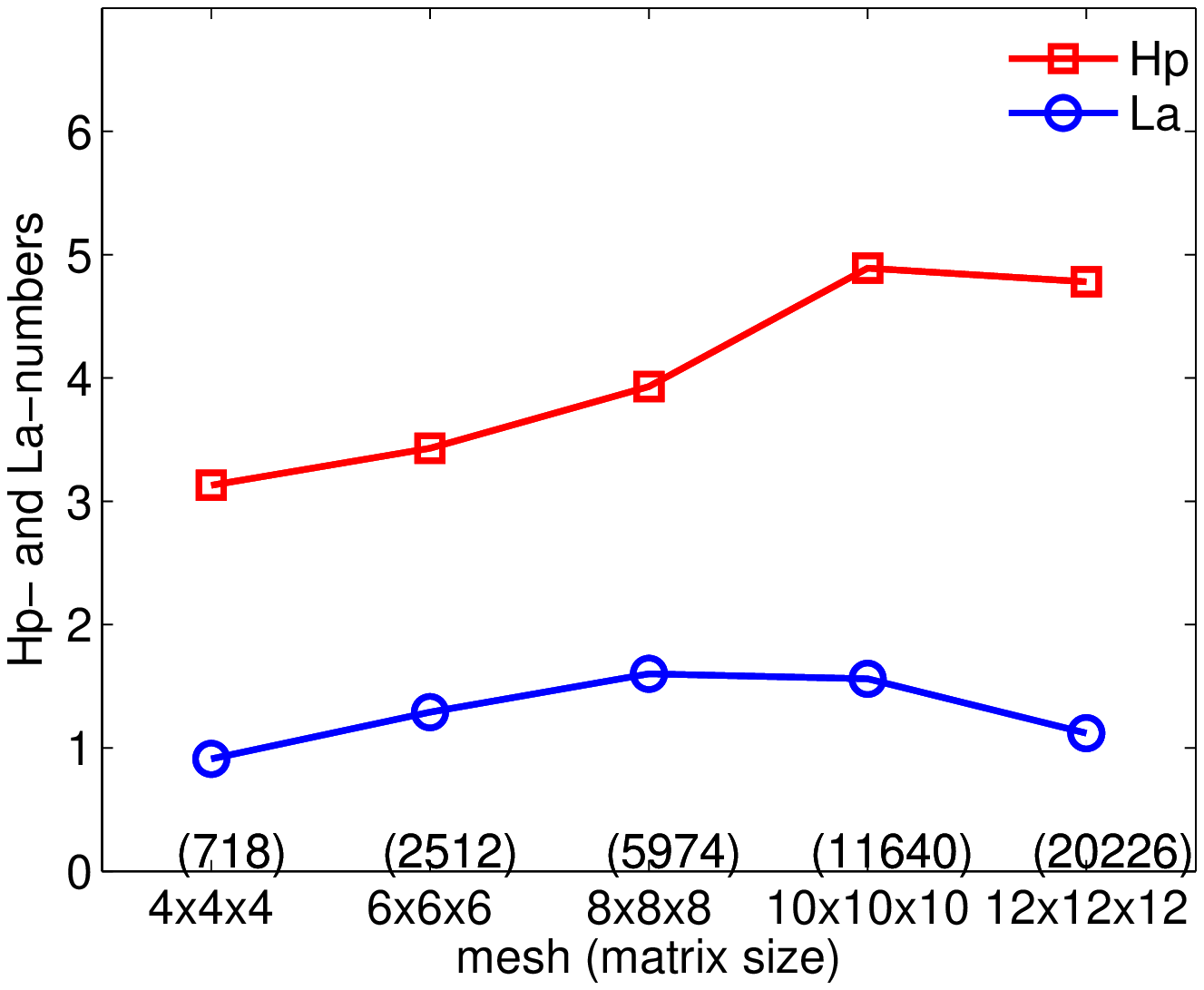}
\includegraphics[width=0.48\textwidth]{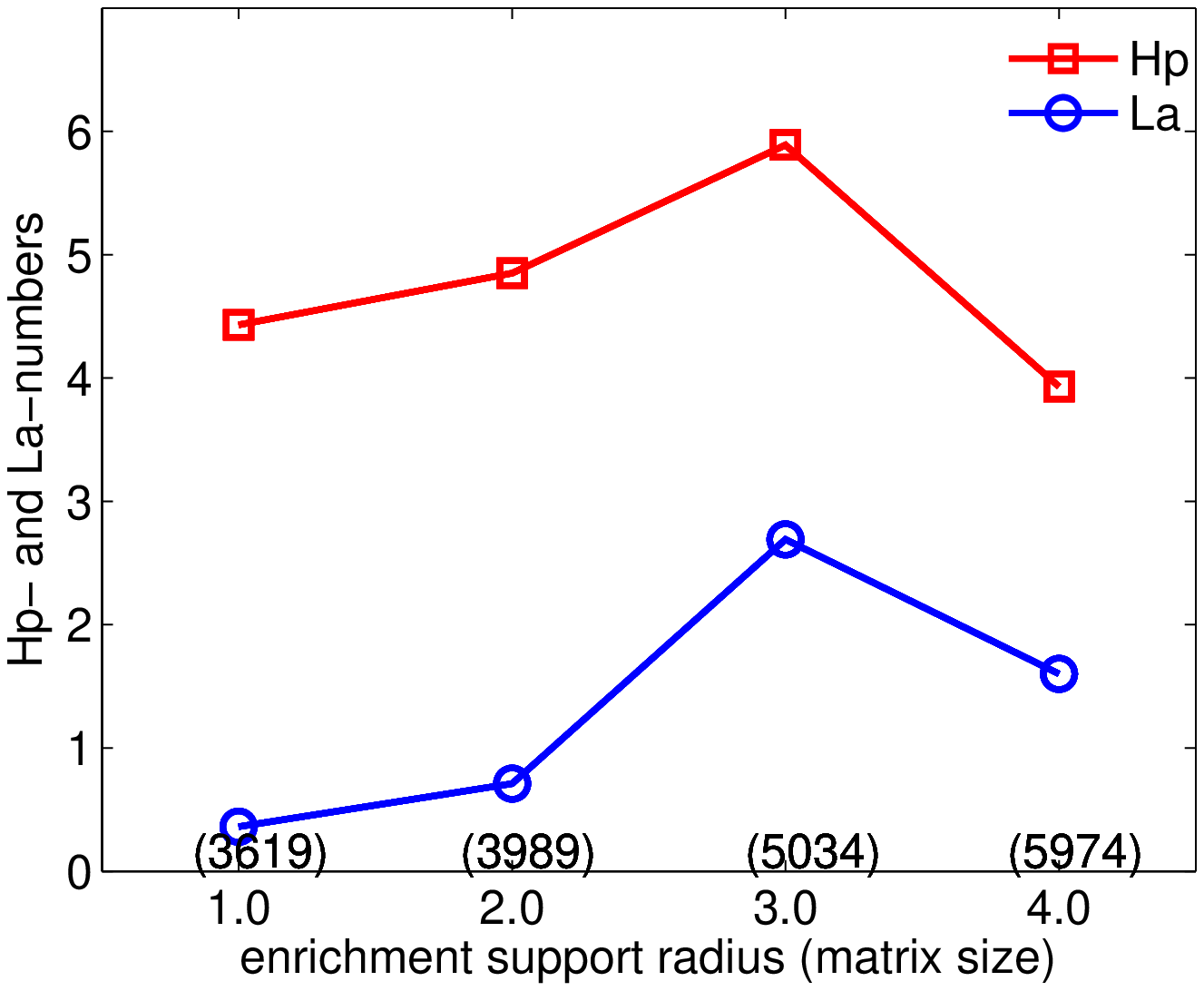}
\caption{CuAl simulation: Hp- and La- numbers (see text) for 
the LABPSD solver for a series of
$n_0 \times n_0 \times n_0$ FE meshes (left) and
enrichment support radii $r_e$ (right).
} \label{fig:cual_ni}
\end{figure}

\paragraph{CuAl} 
The left plot of Figure~\ref{fig:cual_ni}
shows the ${\rm Hp}$- and ${\rm La}$-numbers
for LABPSD for a sequence of refined FE meshes with $r_e = 4$ fixed.
The right plot is for different enrichment support radii $r_e$
and fixed $8 \times 8 \times 8$ FE mesh.
This constitutes a severe test of robustness with respect to 
ill-conditioning since as either the mesh or support radius 
are increased, the conditioning of the GHEP worsens dramatically, 
as shown in Table~\ref{tab:cond}.
In all cases, the rank of $H^{(\rm nl)}$ is $k=19$
and the number of eigenpairs computed per SCF iteration is $m = 10$.

\begin{figure}
\centering
\includegraphics[width=0.48\textwidth]{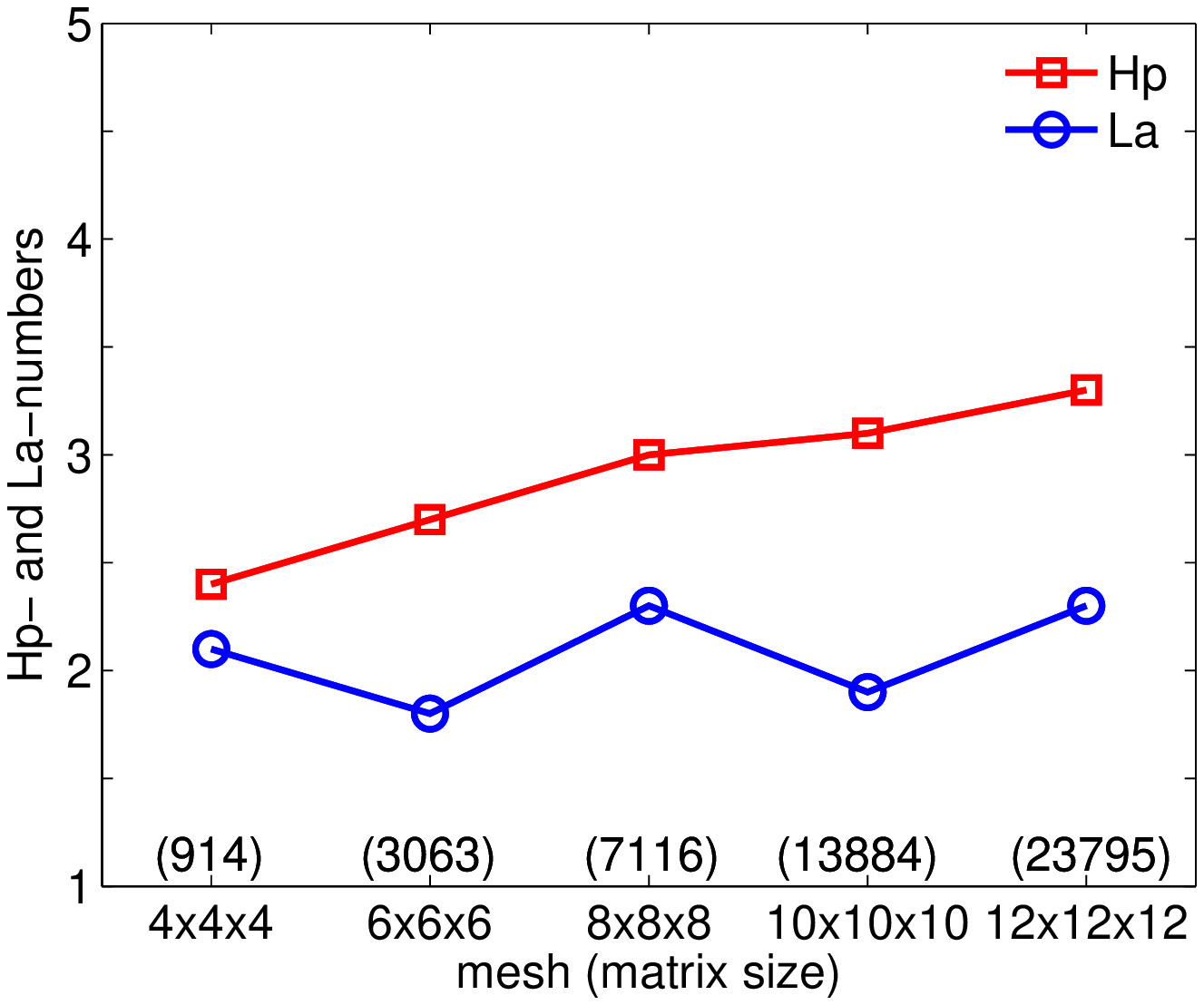}
\includegraphics[width=0.48\textwidth]{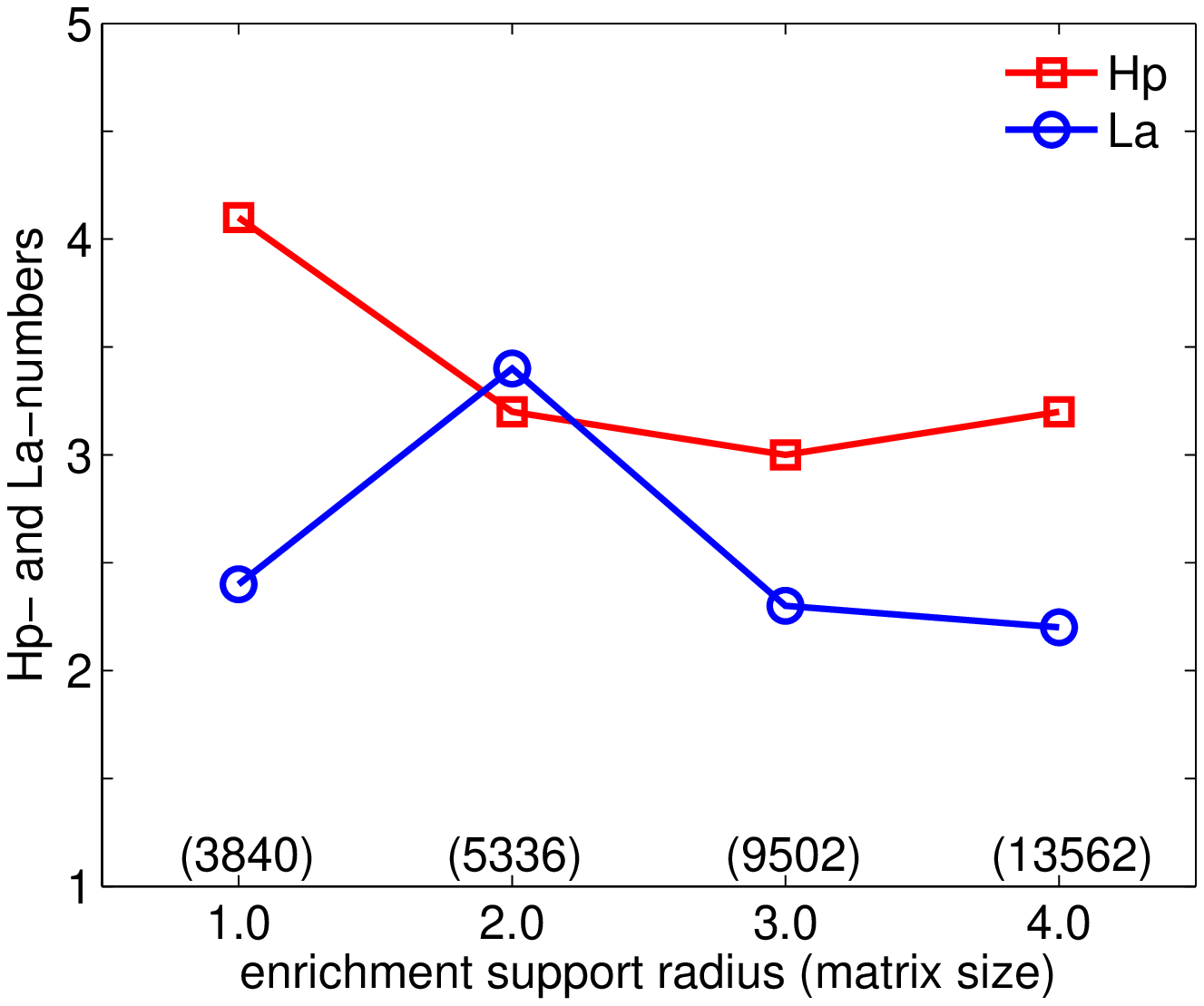}
\caption{CeAl simulation, Hp- and La- numbers (see text) for the 
LABPSD solver for a series of 
$n_0 \times n_0 \times n_0$ FE meshes (left) and 
enrichment support radii $r_e$ (right). 
} \label{fig:ni}  
\end{figure}

\paragraph{CeAl} Similarly, for the CeAl system, 
the left plot of Figure~\ref{fig:ni}
shows the ${\rm Hp}$- and ${\rm La}$-numbers with $r_e = 2.5$.
The right plot is for different enrichment support radii $r_e$  
with the fixed $8 \times 8 \times 8$ FE mesh. 
This constitutes a severe test of robustness with respect to 
ill-conditioning.
In this case, the rank of $H^{(\rm nl)}$ is $k=26$
and the number of eigenpairs computed per SCF iteration is $m = 22$.  

\bigskip 

For both CuAl and CeAl simulations, 
as the mesh is refined or $r_e$ is increased, 
the error of the computed PUFE total energy 
decreases to 
$10^{-6}$ Hartree/atom relative to  the well-converged planewave 
reference. 
Significantly, we observe that {\em all Hp-numbers 
are between 2 and 6}, with only mild dependence on conditioning as it worsens considerably with increasing mesh and support radius. 
Meanwhile, {\em all La-numbers are between 0 and 4}, 
with no apparent dependence on conditioning. 
As we show below (Section \ref{sec:globlochyb}), this is 
in stark contrast to typical global-only or local-only preconditioning 
schemes, which are highly sensitive to the conditioning of the problem.
Furthermore, these Hp- and La-numbers 
are comparable to the typical numbers 
of inner and outer iterations required by the LOBPCG method 
on the well-conditioned \emph{standard} eigenvalue problems 
produced by the planewave \tit{ab initio} method \cite{bott:08}.
This indicates that LABPSD is an efficient method for the
rapid iterative diagonalization of ill-conditioned GHEPs 
produced by nonorthogonal atomic-orbital based methods such as PUFE.

\subsection{Timing}
We now consider the timing of key steps of LABPSD for increasing 
numbers $m$ of eigenpairs. For these purposes, we now focus on the more computationally intensive CeAl system, where 
the dimension of the GHEPs \eqref{ghep} is 
$n_{\rm dof} = 7 \times 12^3 + 11699 = 23795$.  The enrichment 
support radius is $r_e = 2.5$. The rank of $H^{(\rm nl)}$ is $k=26$.

\begin{figure}
\centering
\includegraphics[width=0.7\textwidth]{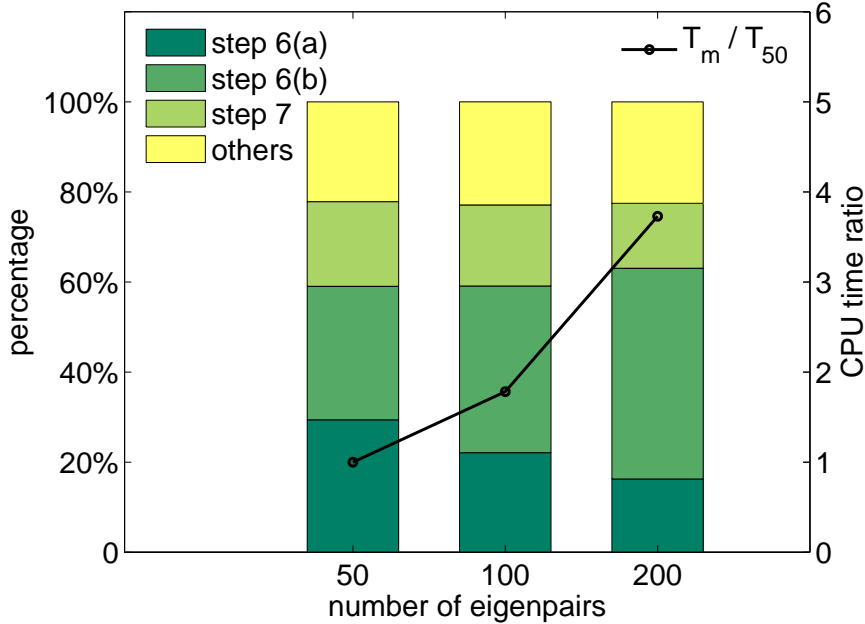}
\caption{Normalized CPU time and percentages 
with increasing $m$ in the CeAl simulation.}\label{fig:timeprofile}
\end{figure}

Figure~\ref{fig:timeprofile} shows the CPU time normalized 
with respect to the CPU time for computing $m=50$ eigenvalues, and
the most time consuming parts for LABPSD are shown for a series of 
PUFE calculations with increasing numbers of eigenpairs $m = 50, 100, 200$
with $m_0 = 10$. 
Each calculation takes 23 SCF iterations to converge to the required tolerance. 
As expected, the CPU time is dominated by the 
preconditioning step 6 at about 60\% of the total time in all cases. 
The cost of the global preconditioner in step 6(a) 
increases as $m$ increases, as expected. However, as a percentage 
of the total, the cost actually decreases, which is a consequence of the 
fact that the cost of the sparse factorization~\eqref{eq:ldl} and 
application of the global preconditioner
is {\em amortized} when more eigenpairs are computed.
On the other hand, the cost of the locally accelerated preconditioners 
in step 6(b) 
increases as a percentage of the total as more eigenpairs are computed. 
The cost of matrix-vector products in step 7 is proportionally  
increased with the number of computed eigenpairs; however,
the overall cost 
is reduced from 20\% to about 15\% of the total when more eigenpairs are computed. 
The costs of all other steps, such as 
setting up the reduced GHEP (step 8),
updating (step 11), and solving the reduced eigenvalue problem (step 9) 
are relatively small at 20\% of the total. 
As $m$ is increased further, 
the solution of the reduced problem must dominate at 
some point due to its $m^3$ scaling. However, at the present 
system sizes, it remains a small fraction of the total.
Overall, when LABPSD is used for computing 4 times more
eigenpairs, namely from $m = 50$ to $m = 200$, the total CPU time 
is also increased by about a factor of 4 (3.73).

We note that the $LDL^\mathrm{H}$ factorization \eqref{eq:ldl} is  
computed only once at the beginning of the SCF cycle.
The CPU time of the factorization \eqref{eq:ldl} is a small percentage 
of the total. Specifically, the $LDL^\mathrm{H}$ factorizations for the two 
$\vm{k}$-points
take just 3\% of the total time when $m=50$, and 0.7\% when $m=200$.

\subsection{Global, local, and hybrid preconditioning}
\label{sec:globlochyb}
Here we compare the hybrid preconditioning scheme to current 
state-of-the-art global preconditioning as in~\cite{blah:10} 
and local preconditioning as in~\cite{pask:05a,ande:05,rays:08}.
Having demonstrated in Sections \ref{eg1} and \ref{sec:iterations} the robustness of the hybrid preconditioner with respect to both the distribution (clustered and nonclustered) and width (hard and soft potentials) of the spectrum, we shall restrict focus here to the more computationally intensive CeAl system, 
where the dimension of the GHEPs is $n=7\times 8^3+3532= 7116$.
The enrichment support radius $r_e = 2.5$, 
the rank of $H^{\rm (nl)}$ is $k=26$, and 
$m = 22$ eigenpairs are computed at each SCF iteration with $m_0 = 3$. 

\begin{figure}
\centering
\includegraphics[width=0.7\textwidth]{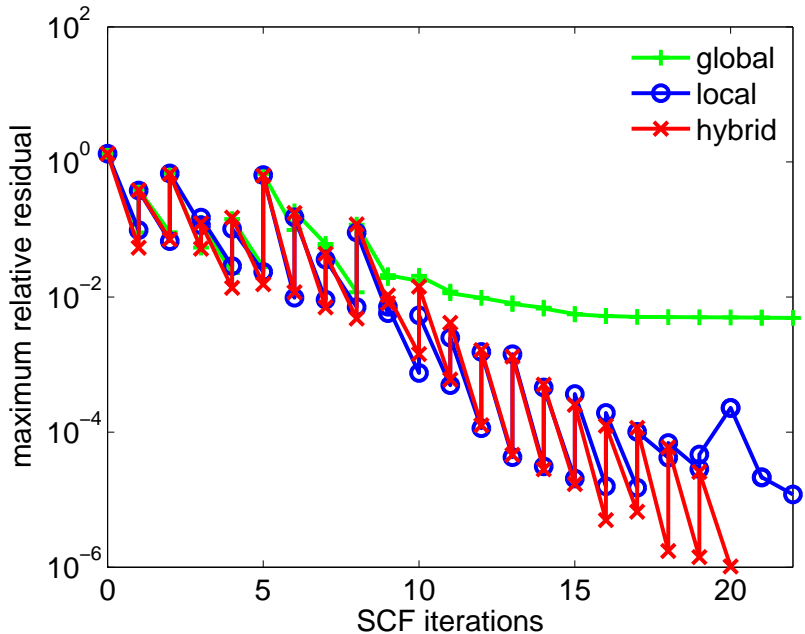}
\caption{
Maximum relative residual norms of GHEPs at the beginning and end
of each SCF iteration, using global, local, and hybrid preconditioners
in the CeAl simulation.
} \label{fig:3precond}
\end{figure}

Figure \ref{fig:3precond} shows 
the maximum relative residual norms of the eigenpairs 
in successive SCF iterations when solving the sequence of GHEPs 
by BPSD with global, local, and hybrid preconditioners.

If we use the global preconditioner step 6(a) only (i.e., without step 6(b)), 
the SCF convergence stagnates after about 9 SCF iterations due to the 
inability of the eigensolver to reduce residuals sufficiently within 
the maximum 200 BPSD iterations. The total CPU time was 11.4 hours, 
due to the relative ineffectiveness of the global preconditioner and 
consequent large number of outer (BPSD) iterations.

On the other hand, if we apply the local preconditioner step 6(b) only, 
without the global preconditioner 6(a), the SCF convergence 
stagnates after about 17 SCF iterations, again due to the inability 
of the eigensolver to reduce the residuals sufficiently 
even with the maximum 100 BPSD and 500 MINRES iterations.\footnote{
We use the locally accelerated
preconditioners after the approximate eigenpairs are localized 
at the 9th SCF iteration. For the first 8 SCF iterations, we
apply the global preconditioner.} 
Due to the large number of both inner (MINRES) and outer (BPSD) iterations,
the total CPU time was 138.6 hours.

In stark contrast, the SCF iteration converges
steadily to the specified tolerance with 
the hybrid preconditioning scheme.
The ${\rm Hp}$- and ${\rm La}$-numbers are 3.0 and 2.3, respectively, 
while achieving smooth SCF convergence at a rate comparable to 
exact diagonalization at each SCF step.   
Due to the small number of both inner and outer iterations, 
the total CPU time was reduced to just 1.3 hours.

\section{Conclusions}\label{sec:conclusion}

We proposed a block hybrid-preconditioned steepest descent 
method, LABPSD, for 
the iterative diagonalization of the sequence of ill-conditioned 
generalized Hermitian eigenvalue problems which arise in 
electronic structure calculations 
using orbital-based nonorthogonal basis sets. 
For such problems, 
the hybrid scheme overcomes the drawbacks of stagnation of
global preconditioners 
and excessive cost of locally accelerated iterative preconditioners. 
PUFE pseudopotential density-functional calculations of CuAl, with soft potentials and degenerate eigenvalues, and CeAl, with hard potentials and nondegenerate spectrum, 
showed Hp- and La-numbers 
comparable to the typical numbers 
of inner and outer iterations required by the LOBPCG method 
on well-conditioned \emph{standard} eigenvalue problems 
produced by the planewave \tit{ab initio} method.
Given the generality of the method and robustness with respect to spectral structure, 
it is expected that the LABPSD method will provide similar benefits to other orbital-based, nonorthogonal electronic structure methods as well. 
Indeed, it is reasonable to expect benefits not only for pseudopotential based methods, as demonstrated here,
 but for  all-electron methods such as APW+lo \cite{sing:05} and LMTO \cite{Skr84} also, since these require diagonalization for just valence states as well (the core states having been solved separately in a spherical approximation).

The LABPSD algorithm and implementation present many opportunities for
future work.  First, similar to \cite{blah:10}, we expect that
the sparse $LDL^{\rm H}$ factorization \eqref{eq:ldl} 
in {\em single precision} or even an incomplete factorization
might be sufficient. This will substantially reduce 
memory and I/O costs for very large systems. 
Secondly, instead of using MINRES for the iterative refinement
in applying locally accelerated preconditioners, 
one can use a simple first-order one-step
iterative method \cite{axel:94}:
\[
\widehat{p}^{(\ell+1)}_i 
= \widehat{p}^{(\ell)}_i 
- \alpha \left[(H^{(i_s)}- \widehat{\varepsilon}_i S) \widehat{p}^{(\ell)}_i - r_i \right]
\] 
with initial $\widehat{p}^{(0)}_i$ from the global preconditioner,  
where $\alpha$ is chosen to minimize the residual norm
of the linear system~\eqref{eq:idealp}.
Our preliminary results are very encouraging, which is
particularly promising for parallel distributed computing. 
In addition, although we have not encountered the rank deficiency 
of the subspace matrix $Z$ produced in step 6 of the LABPSD method,
a rank-revealing re-orthogonalization process 
would be necessary for a general-purpose implementation,
such as in the block steepest descent method implemented in 
EA19 of HSL~\cite{ovtc:10}.

\bigskip

\noindent {\bf Acknowledgments.}
We are grateful to anonymous referees
for their careful reading of the manuscript and most valuable comments. 


\end{document}